\documentclass[aps,pre,preprint,showpacs,floatfix,showkeys]{revtex4-1}
\usepackage{amsmath,bm,epsfig}
\usepackage{graphicx}
\usepackage{latexsym}
\usepackage[section] {placeins}
\usepackage{amssymb}
\usepackage{color}
\usepackage{morefloats}
\usepackage{subfigure}
\begin{document}

\title{Conflicting attachment and the growth of bipartite networks}
\author{Chung Yin (Joey) Leung}
\email[E-mail: ]{cyleung2001@gatech.edu}
\author{Joshua S. Weitz}
\email[E-mail: ]{jsweitz@gatech.edu}
\affiliation{School of Biology, Georgia Institute of Technology, Atlanta, Georgia 30332, USA}
\affiliation{School of Physics, Georgia Institute of Technology, Atlanta, Georgia 30332, USA}

\date{\today}
\begin{abstract}
Simple growth mechanisms have been proposed to explain the emergence of seemingly universal network structures. The widely-studied model of preferential attachment assumes that new nodes are more likely to connect to highly connected nodes. Preferential attachment explains the emergence of scale-free degree distributions within complex networks. Yet, it is incompatible with many network systems, particularly bipartite systems in which two distinct types of agents interact. For example, the addition of new links in a host-parasite system corresponds to the infection of hosts by parasites. Increasing connectivity is beneficial to a parasite and detrimental to a host. Therefore, the overall network connectivity is subject to conflicting pressures. Here, we propose a stochastic network growth model of conflicting attachment, inspired by a particular kind of parasite-host interactions: that of viruses interacting with microbial hosts. The mechanism of network growth includes conflicting preferences to network density as well as costs involved in modifying the network connectivity according to these preferences. We find that the resulting networks exhibit realistic patterns commonly observed in empirical data, including the emergence of nestedness, modularity, and nested-modular structures that exhibit both properties. We study the role of conflicting interests in shaping network structure and assess opportunities to incorporate greater realism in linking growth process to pattern in systems governed by antagonistic and mutualistic interactions.

\end{abstract}
\maketitle

\section{Introduction}\label{sec:intro}

Dynamic network models connect microscopic mechanisms of growth to the emergence of macroscopic structure and function.   In iterative growth models, nodes are sequentially added to an initial small network, and the connectivity of each newly added node is determined based on the existing network and a set of predefined rules. For example, the widely-studied model of preferential attachment (PA) proposed by Herbert Simon in the 1950s~\cite{simon_1955},  extended by Derek de Solla Price in the 1970s~\cite{price_1976} and generalized further by Albert Barab{\'a}si and Reka Albert more recently~\cite{barabasi1999,albert2002}, assumes that the probability that a newly added node connects to an existing node is proportional to the total number of connections (or degree) of the existing node.  The PA model reproduces power-law degree distributions observed in a number of complex scale-free networks such as the World Wide Web (WWW) \cite{barabasi2000}, brain functional networks \cite{eguiluz2005} and the distribution of earthquakes \cite{abe2004}. The preferential attachment mechanism has been verified directly in systems ranging from scientific collaboration networks \cite{newman2001,jeong2003} to protein interaction networks of yeast \cite{eisenberg2003}. Over the past ten years, the PA model has been generalized  to account for other emergent features, including clustering and community structure, beyond that expected given a power-law degree distribution \cite{ren2012,hebert2011,zuev2015}.

Pioneering studies in this area have focused on unipartite networks with only one kind of nodes. Recently, increased attention has been given to bipartite networks in which there are two types of nodes with links exclusively between different types of nodes, for example, scientific collaboration networks in which authors are associated with papers they publish \cite{newman2001i,newman2001ii} and ecological networks of species interactions ranging from mutualistic plant-pollinator networks \cite{olesen2006} to antagonistic parasite-host networks \cite{poulin2010}. A number of studies have generalized the PA model to bipartite networks. While PA models of bipartite networks can generate power-law degree distributions in certain situations \cite{zhang2015}, such models also lead to other forms of degree distribution such as an U-shaped distribution with many nodes with few and many nodes with many links \cite{peruani2007} and a shifted power-law distribution \cite{Zhang2013}. 

Ecological networks exhibit complex network properties that are not readily explained by standard PA models or recent generalizations to bipartite networks. A recurrent pattern is that of nestedness, where species with a small number of links (specialists) have a subset of the links of species with a greater number of links (generalists) \cite{bascompte2003}. In other words, specialists interact with generalists while generalists interact with both specialists and generalists. In the limit of perfect nestedness, the degree distribution is uniform rather than scale-free. In addition, ecological networks can also exhibit modularity where the species are separated into groups or modules with species interacting strongly within a module but not between different modules \cite{olesen2007}. These network properties play an important role in the function and stability of ecosystems. For example, it has been suggested that nestedness enhances stability in mutualistic networks \cite{fortuna2006,thebault2010} while modularity promotes stability in food-webs \cite{thebault2010,stouffer2011}. However, other studies reveal that nestedness and modularity are not always mutually exclusive \cite{fortuna2010}, and the relationship between the structure of ecological networks and ecosystem stability largely depends on ecological parameters such as growth rates of the species \cite{jover2015}. Nonetheless, network structural properties such as nestedness and modularity are expected to significantly influence the dynamics of an ecosystem.

For competitive ecosystems, there exists another type of incompatibility with respect to the PA mechanism as there is a conflict of interest between the two types of nodes (e.g. parasite and host) in terms of the preferred level of connectivity. Other examples of conflict of interest in networks include server-hacker networks\cite{debar_1999} and surveillance-rogue networks \cite{krebs_2002,borgatti_2009}. Such conflict can manifest itself in terms of distinct preferences, i.e., nodes of one type ``prefer'' to have more links and nodes of the other type ``prefer" to have fewer links. The interaction between viruses and their microbial hosts is an increasingly well-studied example of competitive ecosystems.  In virus-microbe communities it is beneficial for viruses to infect microbes in order to reproduce, while the microbial hosts benefit by avoiding infection and subsequent lysis; complications can arise in the case of long-term infections of microbes by viruses \cite{feiner2015}. Observations of the selective benefit of increased resistance and infectivity, for microbes and viruses respectively, date back to seminal work by Salvador Luria and Max Delbr\"{u}ck  in the 1940s \cite{luria_1943,luria_1945}. Empirical studies of infection networks between bacteria and phage, or a type of virus that exclusively infects bacteria, have revealed complex patterns of cross-infection (see the review in Weitz \textit{et al.} \cite{weitz2013}) including nestedness \cite{flores2011} and modularity \cite{flores2013}.  Further, in the marine example, the identified modules also had an internal nested structure, i.e., they had a multi-scale interaction network pattern \cite{flores2013}.

In this manuscript, we propose a conflicting attachment model for the growth of bipartite networks inspired by the competitive ecological interactions between viruses and their microbial hosts.  The governing mechanism of the conflicting attachment model is the duplication of nodes, akin to a mutation event, with link rewiring that favors removal and addition of links for the two types of nodes respectively.   We assume that link rewiring comes at a cost, with two parameters governing the relative costs of rewiring for the two types of nodes respectively.   As we show, changing the strengths of trade-offs leads to qualitative changes in network structures consistent with empirical observations, including regimes characterized by elevated nestedness, modularity, multi-scale nested-modular structure, and a superimposition of nested and modular patterns. We also explore how the current model may be used to probe the relationship between microscopic mechanisms and macroscopic patterns in the growth of bipartite networks more generally - including both antagonistic and mutualistic systems.

\section{Model}\label{sec:model}

\subsection{Bipartite network growth model}

We propose a stochastic model of bipartite network growth that incorporates a conflict of interest in its generative mechanism. The state of the system is described in terms of a graph $G\equiv \{H,V,E\}$ where $H$ are the host nodes, $V$ are the virus nodes and $E$ are the links between viruses and hosts. The graph $G$ can be represented by an $m \times n$ incidence matrix $\mathbf{B}$ where $m$ and $n$ is the number of host nodes and virus nodes respectively, and $B_{ij}=1$ if host $i$ is infected by virus $j$, otherwise $B_{ij}=0$.

\begin{figure}[!htbp]
\centering
\includegraphics[width=12cm]{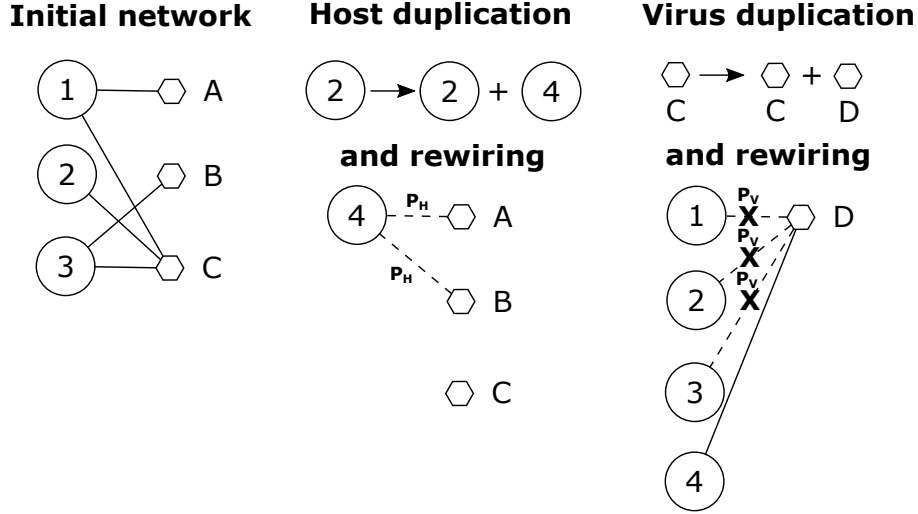}
\caption{Schematics showing selection of nodes for duplication and rewiring where host $2$ and virus $C$ would be selected (left). The duplication and stochastic rewiring of links are also shown for the new host (middle) and virus (right). Solid lines are connections and dashed lines are potential links associated with a rewiring probability that corresponds to an addition and deletion of link for host and virus, respectively.}
\label{fig1:sch}
\end{figure}

Each network growth step consists of two parts, the first is the duplication and rewiring of a particular host and the second is the duplication and rewiring of a particular virus. In the first part of each model step, the host $H_r$ with the smallest degree $k_r=\sum_jB_{rj}$ is selected for duplication and a new host is generated that inherits all the links of $H_r$.   If more than one host has the smallest degree, then one of them is selected at random. The new node is denoted as $H'$, e.g., node $2$ - the most resistant host - would be selected for duplication in the example network in Fig. \ref{fig1:sch}.  The rewiring of $H'$ involves, first, the removal of all links - equivalent to evolving resistance to all viruses that infect $H_r$. Then, new links are added with probability $P_H$ to each virus unconnected to its ancestor $H_r$. This loss of resistance can be understood as the trade-off of developing new resistance.

In the second part of each model step, the virus $V_g$ with the largest degree $d_g=\sum_j B_{jg}$ is selected for duplication and a new virus is generated that inherits all the links of $V_g$.  If more than one of them have the largest degree, then one of the viruses with the largest degree is selected at random. The new virus is denoted as $V'$, e.g., node C - the most generalist virus - would be selected for duplication in the example network in Fig. \ref{fig1:sch}.  The rewiring of $V'$ involves, first, the addition of links to all hosts previously inaccessible by its ancestor $V_g$. Then, existing links of $V_r$ are removed with probability $P_V$. This loss of connectivity can be understood as the trade-off of exploiting new types of hosts.

The conflict of interest is embedded in both the node duplication and the link rewiring processes. Duplication of the most resistant host and most generalist virus presumes that fitness is correlated with the scope of resistance and infectivity, respectively. Similarly, the newly evolved host gains resistance while the newly evolved virus develops new infectivity - leading to decreases and increases in network connectivity.  The cost of such gains are parameterized by  $P_H$ and $P_V$. When these parameters approach zero,  then there is little cost in developing new resistance or infecting new host types. In contrast, when these probabilities approach one, then developing resistance or gaining access to new hosts involves the loss of resistance or the loss of existing infectivity, for hosts and viruses respectively. 

The bipartite growth process is intentionally set to be unbounded, similar to that of the original PA model.  Such unbounded growth can be interpreted as akin to the accumulation of strain types in coevolution experiment of phage and bacteria, for which cross-infectivity can then be assessed within and across time-points (e.g., ~\cite{poullain_2008,scanlan2011,meyer_2012repeatability,brockhurst2013,sieber_isme2014}).

\subsection{Quantifying bipartite network structure}

We are interested in the modularity and nestedness of the networks generated and employ standard metrics to quantitatively measure these structural properties. The modularity of a network is defined as the ratio of links within each module subtracted by the ratio expected to result from randomness \cite{newman2006}, and for a bipartite network it is given by \cite{barber2007}
\begin{equation}\label{eq:modu}
Q\equiv \frac{1}{E}\sum_{i,j}(B_{ij}-\frac{k_id_j}{E})\delta(g_i,g_j)
\end{equation}
where $E$ is the total number of links in the network, $k_i\equiv\sum_jB_{ij}$ is the degree of node $H_i$ and $d_j\equiv\sum_iB_{ij}$ is the degree of $V_j$.

The nestedness of a network can be quantified by the nestedness metric based on overlap and decreasing fill (NODF) \cite{almeida2008}. This metric is based on two observations. Firstly, in a perfectly nested network, the nodes range from generalist to specialist (decreasing fill), therefore their degrees are never identical. Secondly, in such a network the interactions of a specialist (lower degree) always form a subset of the interactions of a generalist (higher degree). The NODF metric therefore measures the nestedness among the rows (hosts $H$) of the incidence matrix by defining a matrix $M^H_{ij}$ for each pair of rows such that
\begin{equation}\label{eq:MHij}
M^H_{ij}\equiv
\begin{cases}
0 & \text{if } k_i=k_j \\
\sum_k B_{ik}B_{jk}/min(k_i,k_j) & \text{if } k_i\neq k_j
\end{cases}
\end{equation}
where $k_i$ and $k_j$ are the degrees of hosts $i$ and $j$ respectively. Suppose $k_i\geq k_j$, $M^H_{ij}$ is equal to $1$ when the links of host $j$ form a subset of the links of host $i$, less than $1$ if only some of the links of host $j$ overlap with links of host $i$, and equals $0$ if $k_i=k_j$. $M^H_{ij}$ is therefore maximized when the host connections satisfy the criteria of a perfectly nested network. A matrix $M^V_{ij}$ is defined similarly for the nestedness across the columns (viruses), and the overall nestedness of the network is given by 
\begin{equation}\label{eq:nest}
{\mathcal{N}}_{NODF}\equiv\frac{\sum_{i<j}(M^H_{ij}+M^V_{ij})}{m(m-1)/2+n(n-1)/2}.
\end{equation}

The NODF metric of nestedness is independent of the ordering of the nodes and gives a single value for each network. On the other hand, the calculation of modularity depends on how the network is partitioned. To identify the partition that maximizes modularity, we use the Bipartite Recursively Induced (BRIM) method~\cite{barber2007,newman2006}. This algorithm detects modules by maximizing $Q$ iteratively, with the nodes assigned to sets that recursively draw each other into modular structures. The partition of nodes thus identified is guaranteed to be a local maximum of $Q$. Computations of the modularity and nestedness, and visualization of the bipartite networks are carried out using the BiMat MATLAB package \cite{flores2014}.

For each set of parameters, the modularity and nestedness are calculated for each network in an ensemble of networks, and the average value of the modularity and nestedness are then determined. To ascertain the statistical significance of the modularity and nestedness values, they can be compared to the distribution of values generated via a null model.

\section{Results}\label{sec:results}

\subsection{Emergent network structure given conflicting growth mechanisms}
\begin{figure}[!htbp]
\centering
\subfigure[]{\includegraphics[width=8cm]{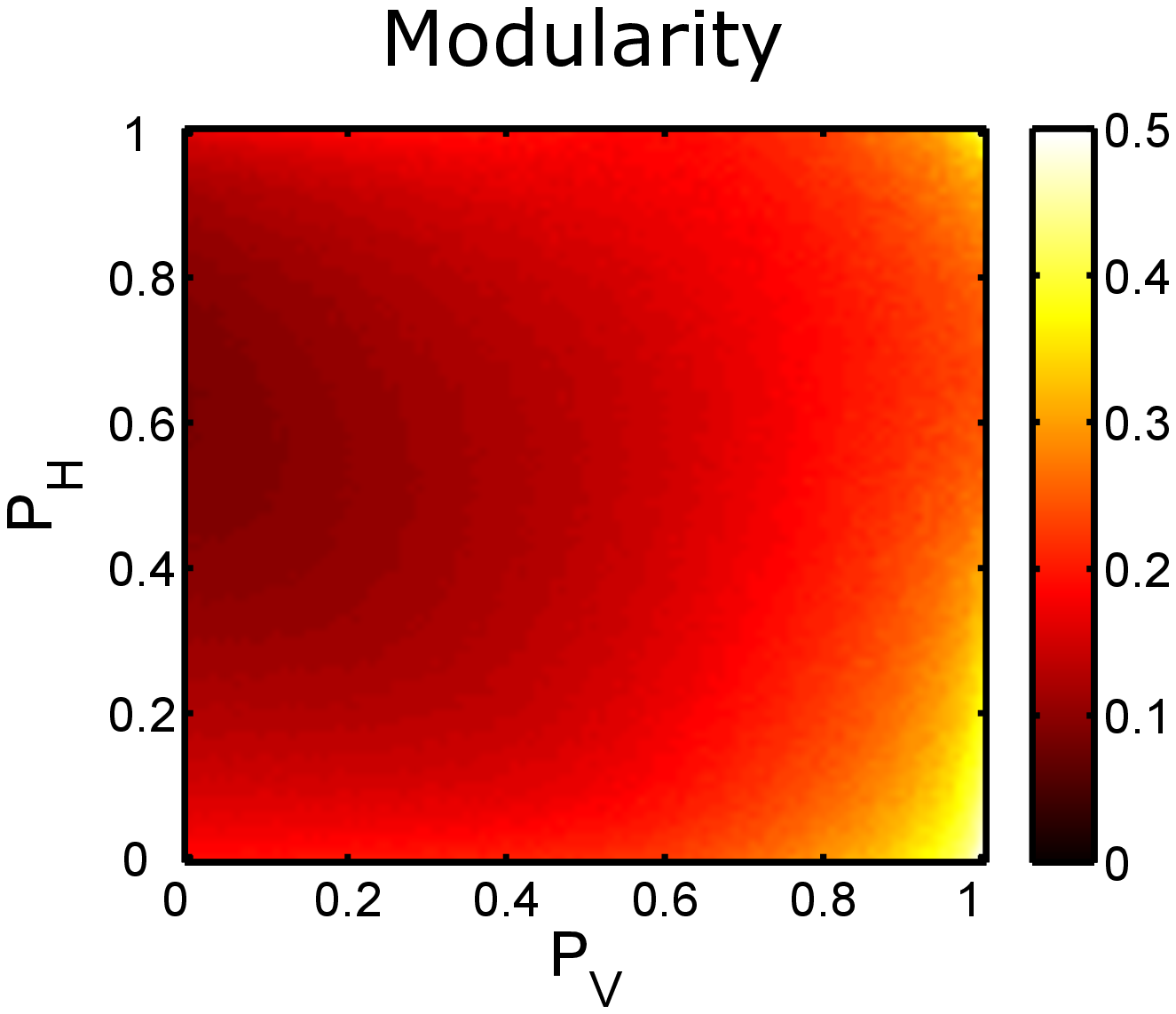}}
\subfigure[]{\includegraphics[width=8cm]{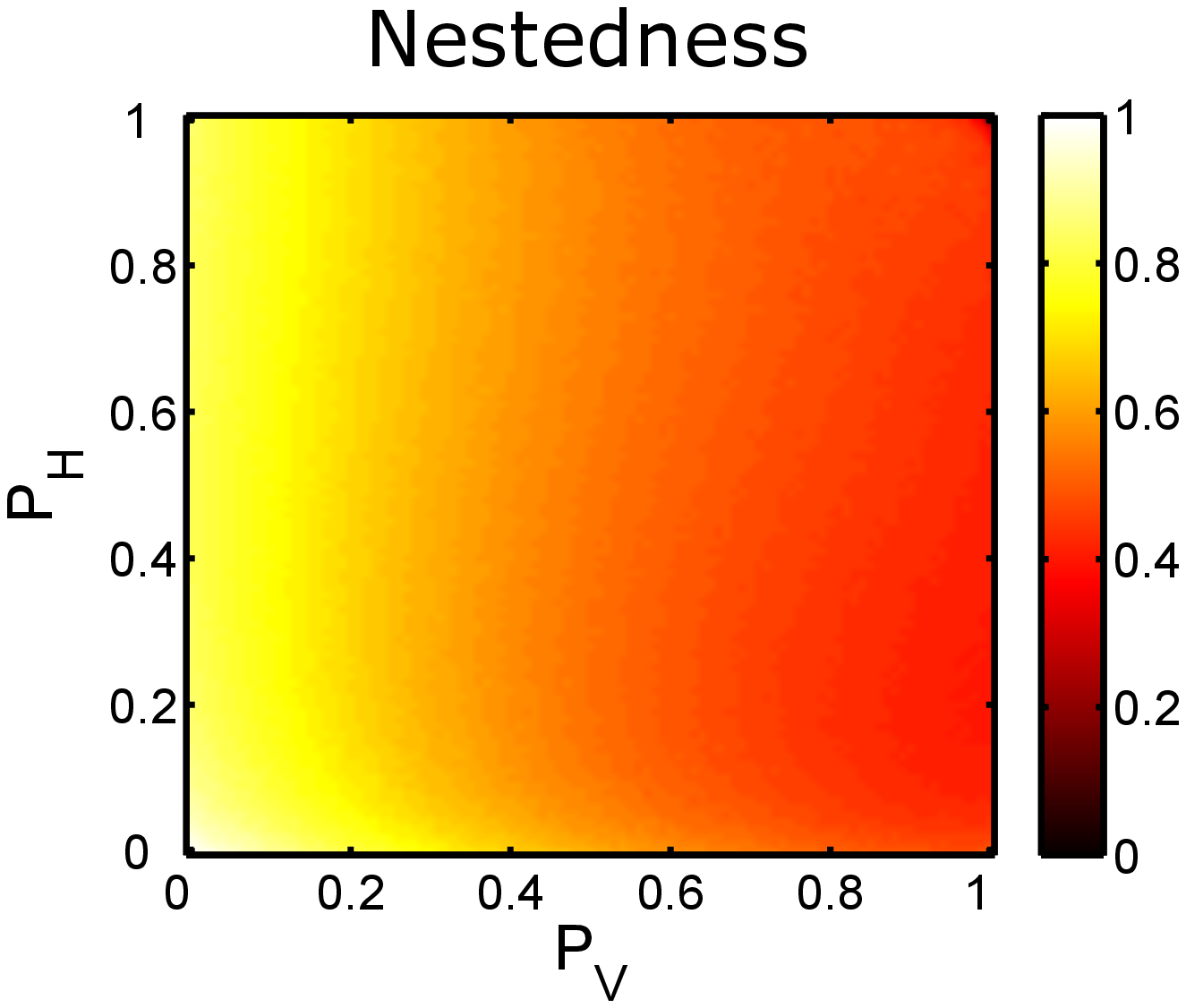}}
\caption{(Color online) Heat map showing the dependence of (a) modularity $Q$ and (b) nestedness ${\mathcal{N}}_{NODF}$ on the parameters $P_H$ and $P_V$. The network size is $n=50$ and values are averaged over $20$ networks.}
\label{fig2:phase}
\end{figure}

The dynamics of network growth given conflicting attachment can be quantified in terms of emergent network structures.  Fig.~\ref{fig2:phase} shows variation in the measured nestedness, ${\mathcal{N}}_{NODF}$ and modularity $Q$ of ensembles of networks as a function of the two governing cost parameters: $P_H$ and $P_V$.  In these simulations, nestedness is measured using the NODF metric (see Methods and \cite{almeida2008}) and modularity is measured using Barber's bipartite modularity metric (see Methods and \cite{newman2006,barber2007}). As an illustrative example, we generate networks with $50$ hosts and $50$ viruses, but we subsequently show that there are simple scaling relationships between the network properties and network size, so the qualitative trend also applies to other network sizes. Across the entire parameter range, modularity tends to increase with $P_V$ while nestedness tends to decrease with $P_V$.  One exception to this trend occurs when $P_H=0$ and $P_V=1$, which corresponds to  a regime of high modularity and intermediate nestedness.  

\begin{figure}[!htbp]
\centering
\subfigure[]{\includegraphics[width=3.5cm]{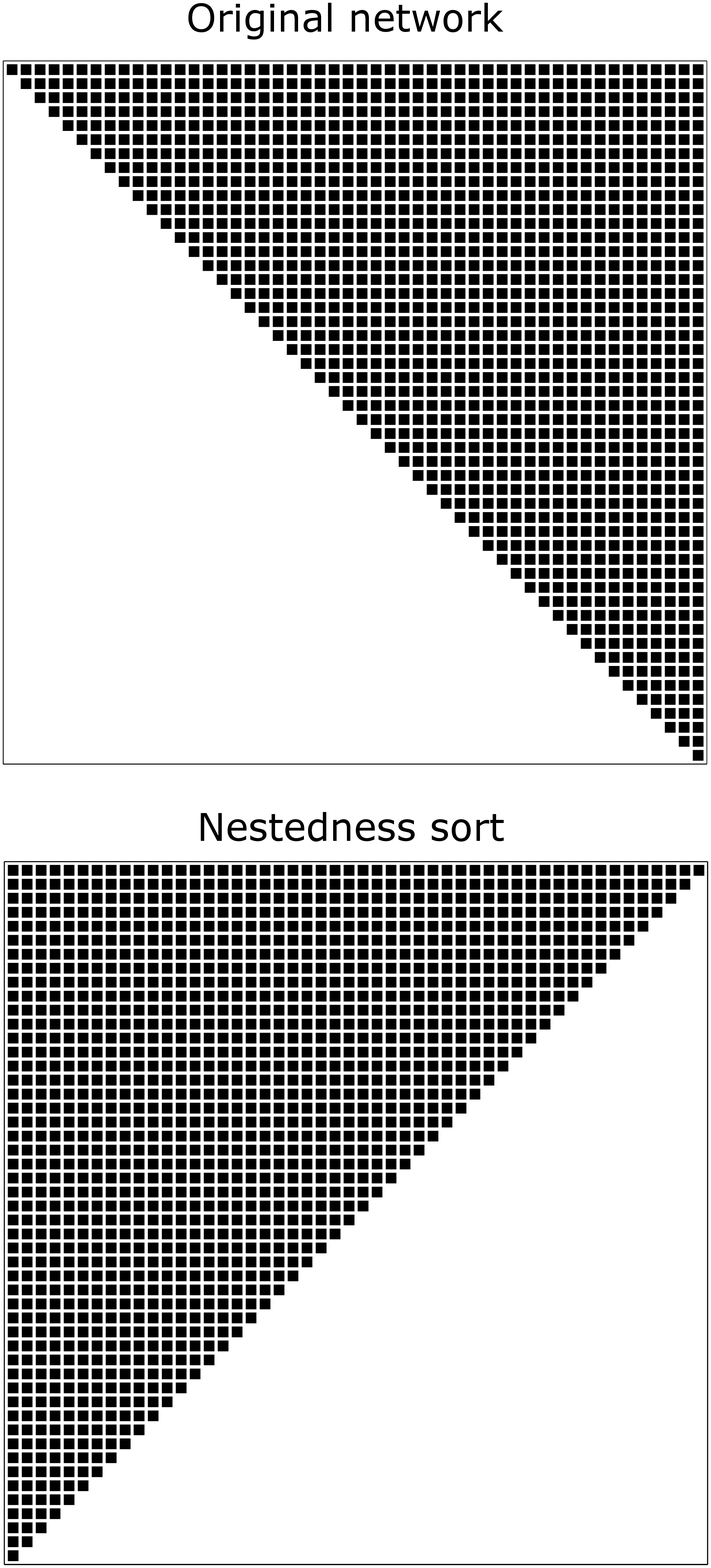}}
\hspace{0.3cm}
\subfigure[]{\includegraphics[width=3.5cm]{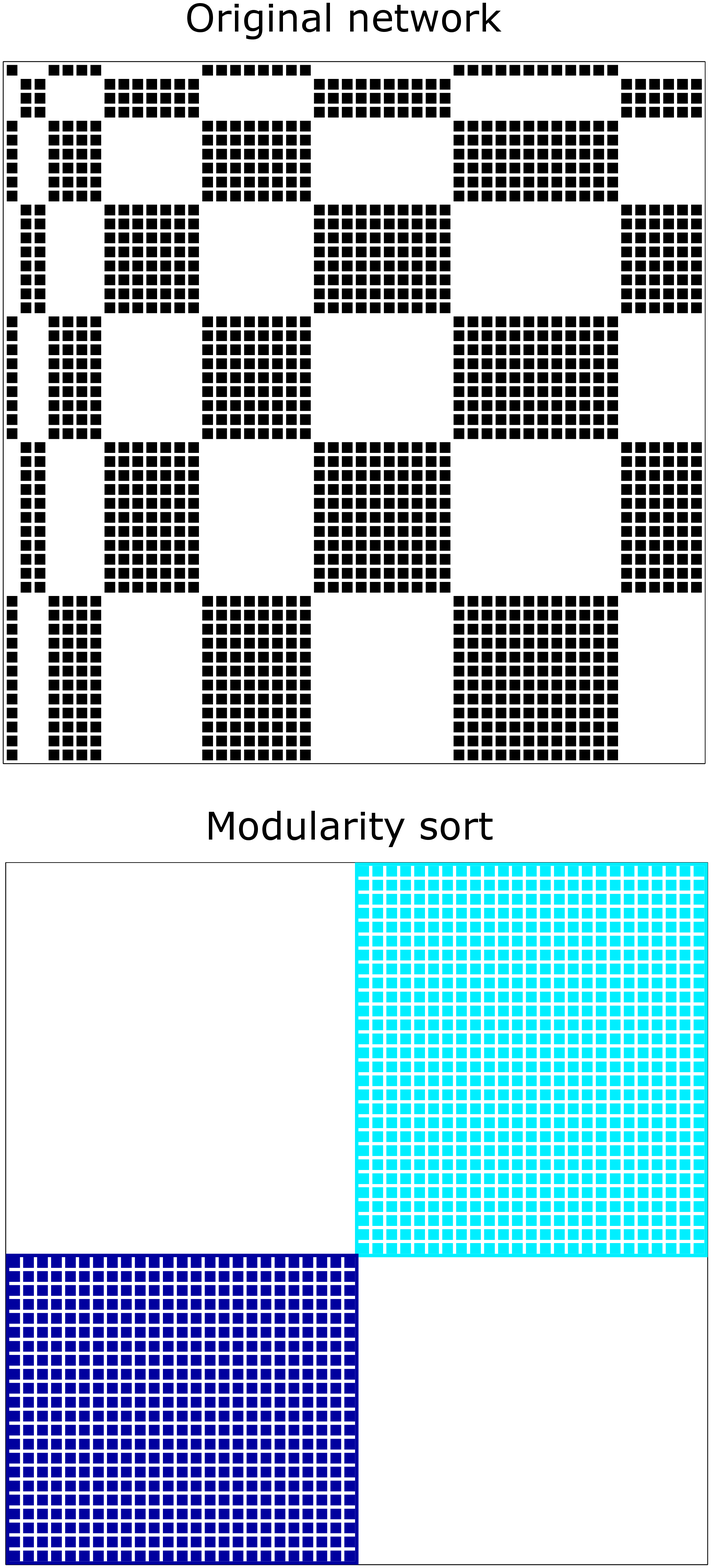}}
\hspace{0.3cm}
\subfigure[]{\includegraphics[width=3.5cm]{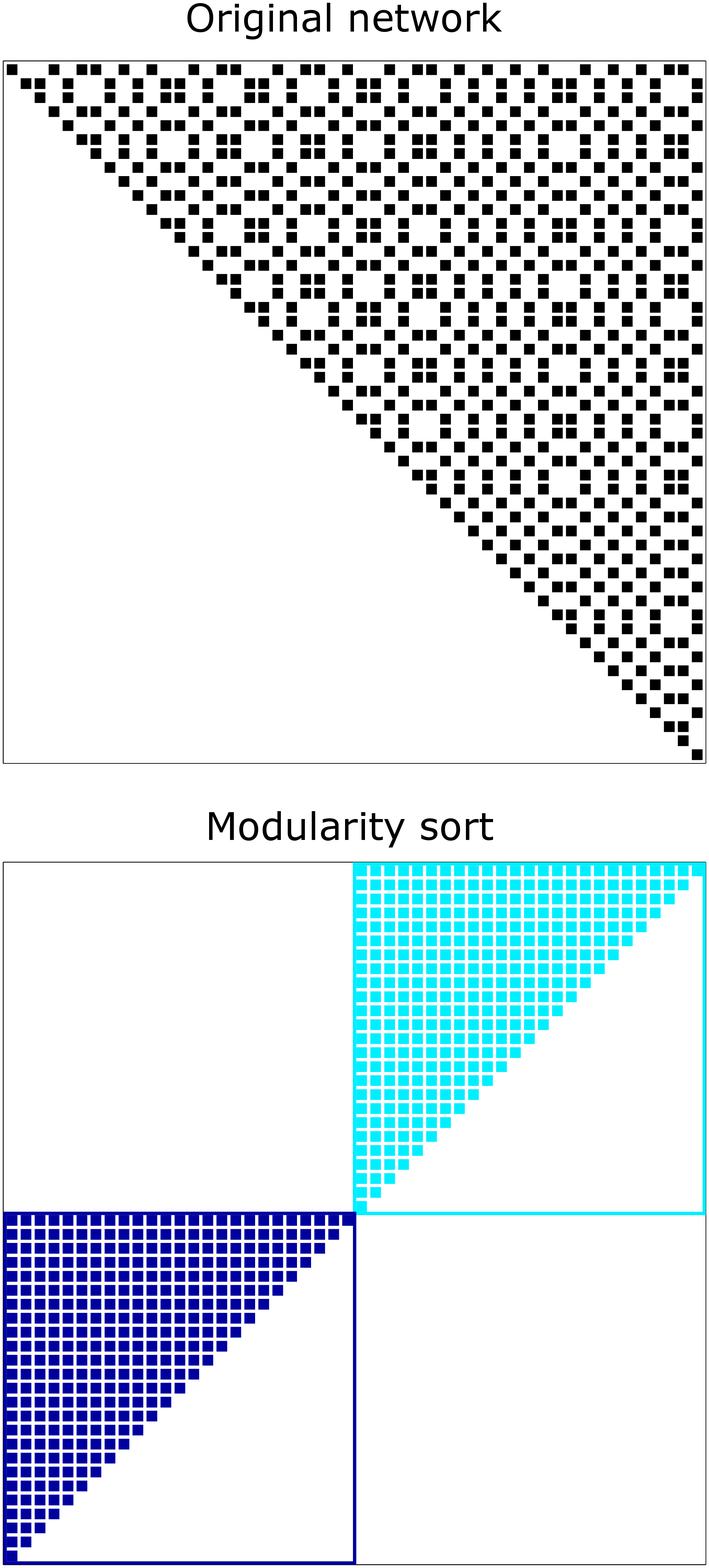}}
\hspace{0.3cm}
\subfigure[]{\includegraphics[width=3.5cm]{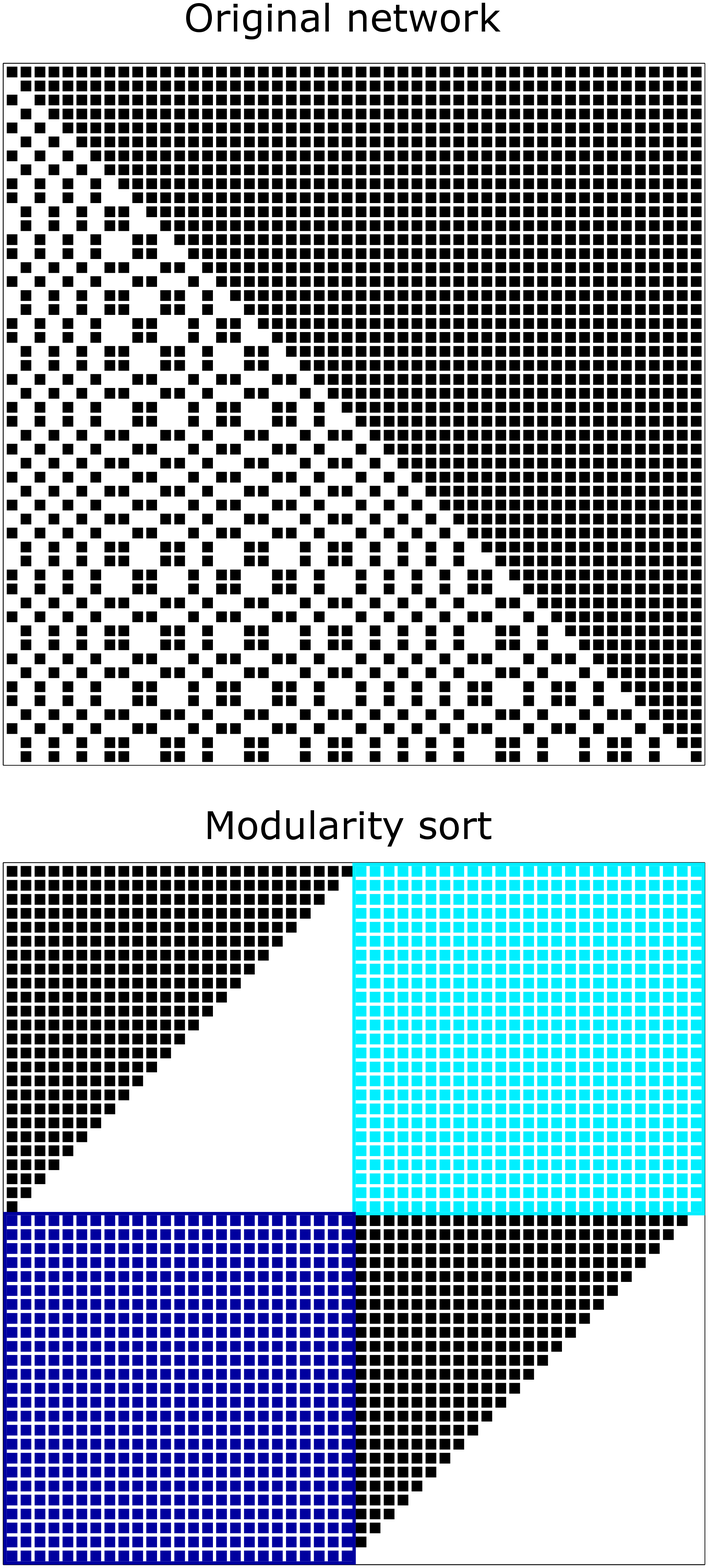}}
\caption{(Color online) Examples of networks with asymptotic values of $P_H$ and $P_V$: (a) $P_H=P_V=0$ (nested) with the original network and a sorted version of the network in the standard nested form with rows and columns in descending node degree, (b) $P_H=P_V=1$ (modular) with the original network and a sorted version highlighting the modularity of the network (c) $P_H=0$, $P_V=1$ (multi-scale nested-modular) with original network and the modularity sorted version, and (d) $P_H=1$, $P_V=0$ (superimposed nested-modular) with original network and modularity sort.}
\label{fig3:net}
\end{figure}

The different asymptotic cases illustrate the variation in emergent network structures.  When $P_H=P_V=0$ there is no cost to developing resistance or acquiring an expanded host range.  In that limit, the growth process generates a perfectly nested network with ${\mathcal{N}}_{NODF}=1$ where the hosts and viruses can be ranked in a hierarchy of strictly increasing node degree and the interactions of the specialists are always a subset of the generalists (Fig.~\ref{fig3:net}(a)). When $P_H=P_V=1$, the costs are maximal for acquiring resistance and infectivity.  In turn, the emergent networks are perfectly modular with two modules wherein all hosts and viruses within a module are fully connected, and there are no links across different modules (Fig.~\ref{fig3:net}(b)). When $P_H=0$ and $P_V=1$, the costs are asymmetric, such that hosts have no cost to acquire resistance and viruses have maximal costs. In this case, the emergent networks have a multi-scale nested-modular structure where the nodes can be separated into two modules with each of the module being a perfectly nested network (Fig.~\ref{fig3:net}(c)). Interestingly, this type of multi-scale nested-modular pattern has been reported in analysis of empirical data of phage-bacteria infection networks \cite{flores2011}. For the opposite case of $P_H=1$ and $P_V=0$ in which host resistance has maximal costs while virus infectivity incurs none, the network shows a superimposed pattern of modularity and nestedness. In such a superimposed nested-modular network, the hosts and viruses can be separated into two modules, with hosts and viruses within the same module fully-connected and the interactions across different modules exhibiting a nested pattern (Fig.~\ref{fig3:net}(d)). Note also that this structure is related to the multi-scale nested-modular structure by swapping all links and nonexistent links ($B_{ij}\rightarrow 1-B_{ij}$). For all of the asymptotic cases with elevated nestedness, including the perfectly nested ($P_H=P_V=0$), multi-scale nested-modular ($P_H=0$, $P_V=1$), and superimposed nested-modular ($P_H=1$, $P_V=0$) case, the degree distributions of the hosts and viruses are uniform with a range defined by the most specialist and the most generalist nodes. The skewness of the distributions are therefore zero. For the perfectly modular case ($P_H=P_V=1$) the degree distribution is bimodal with each peak corresponding to the number of hosts (or viruses) in a module. The symmetry of the distribution also ensures a low skewness. The low skewness indicates that the distributions are not described by power law as expected from preferential attachment. In the next section, we explain the processes by which variation in microscopic mechanisms of rewiring generates distinct classes of network structures.

\subsection{Mechanisms for the emergence of modularity and nestedness given conflicting attachment}

\begin{figure}[!htbp]
\centering
\subfigure[]{\includegraphics[width=2cm]{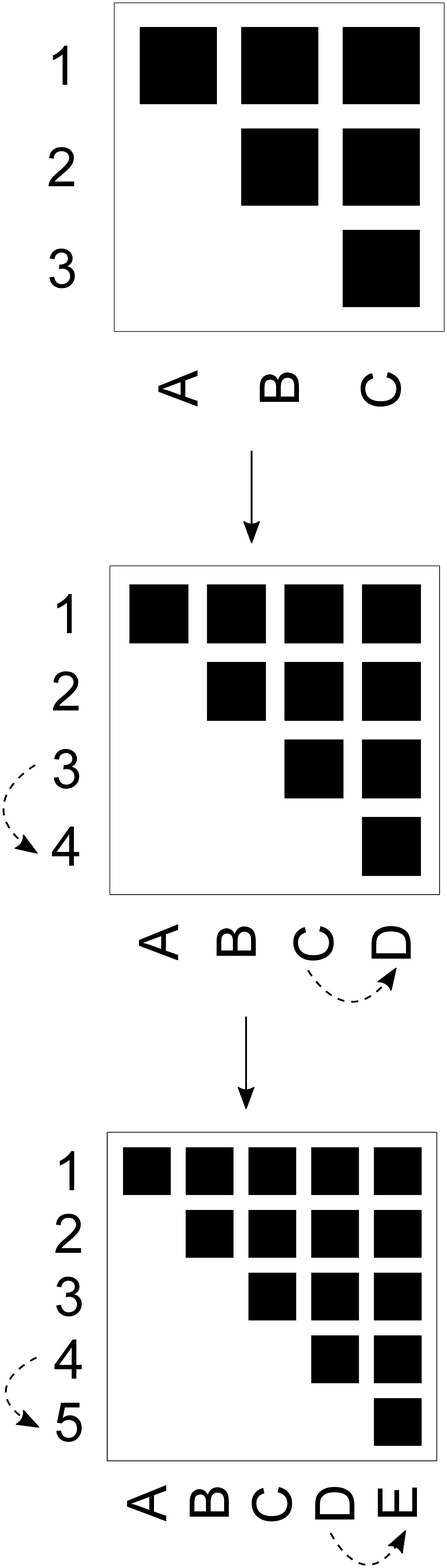}}
\hspace{0.5cm}
\subfigure[]{\includegraphics[width=4cm]{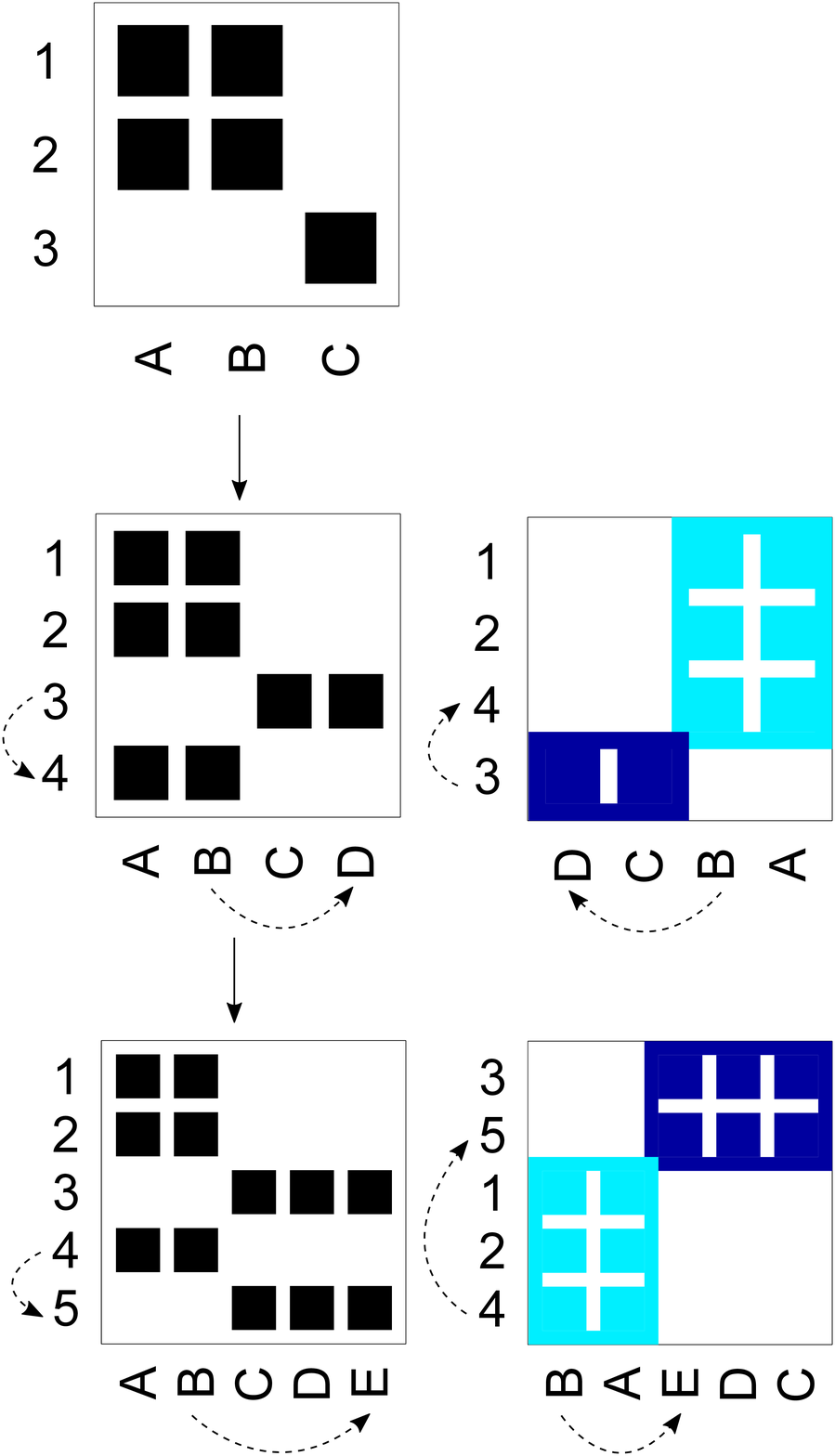}}
\hspace{0.5cm}
\subfigure[]{\includegraphics[width=4cm]{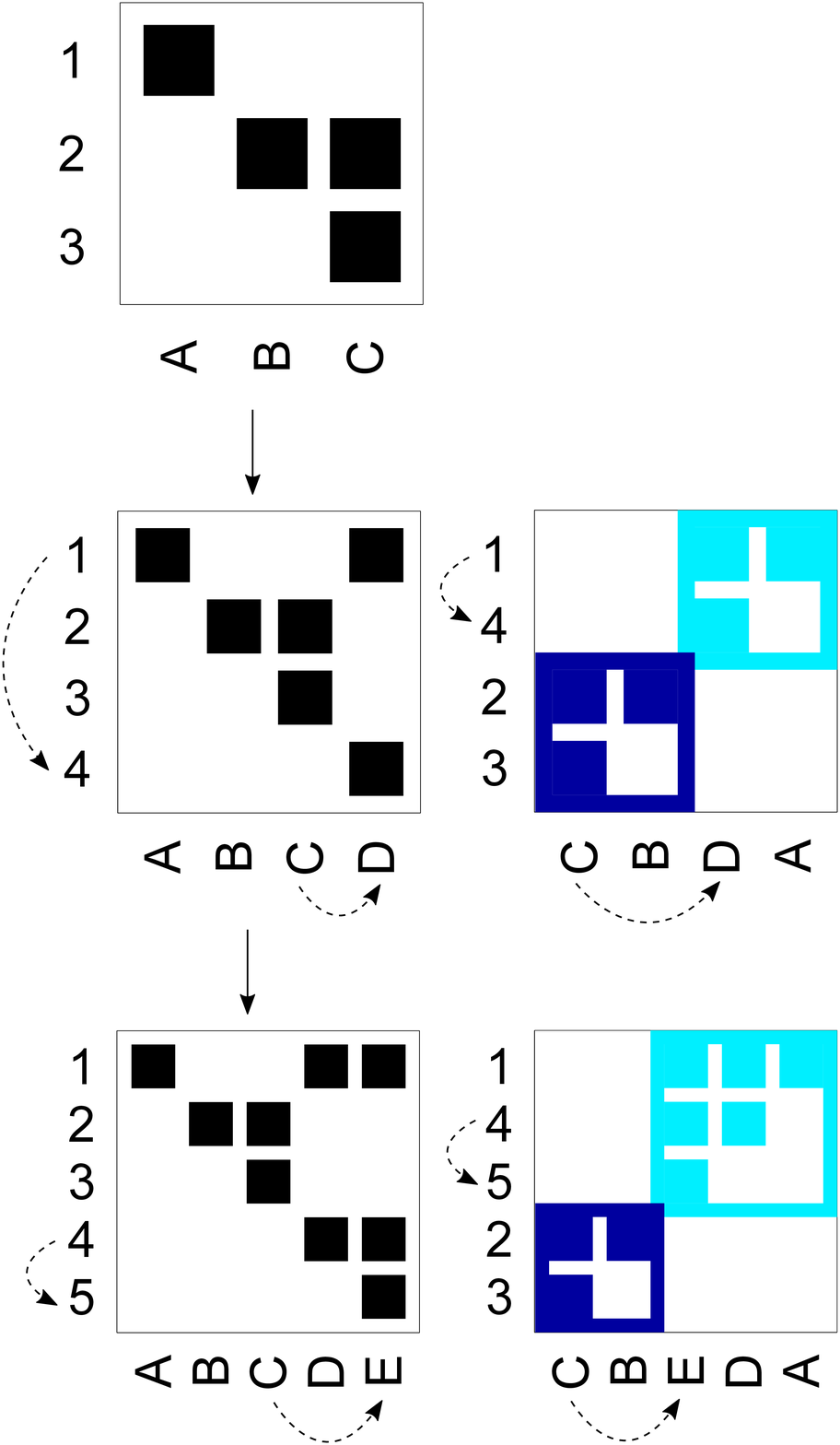}}
\hspace{0.5cm}
\subfigure[]{\includegraphics[width=4cm]{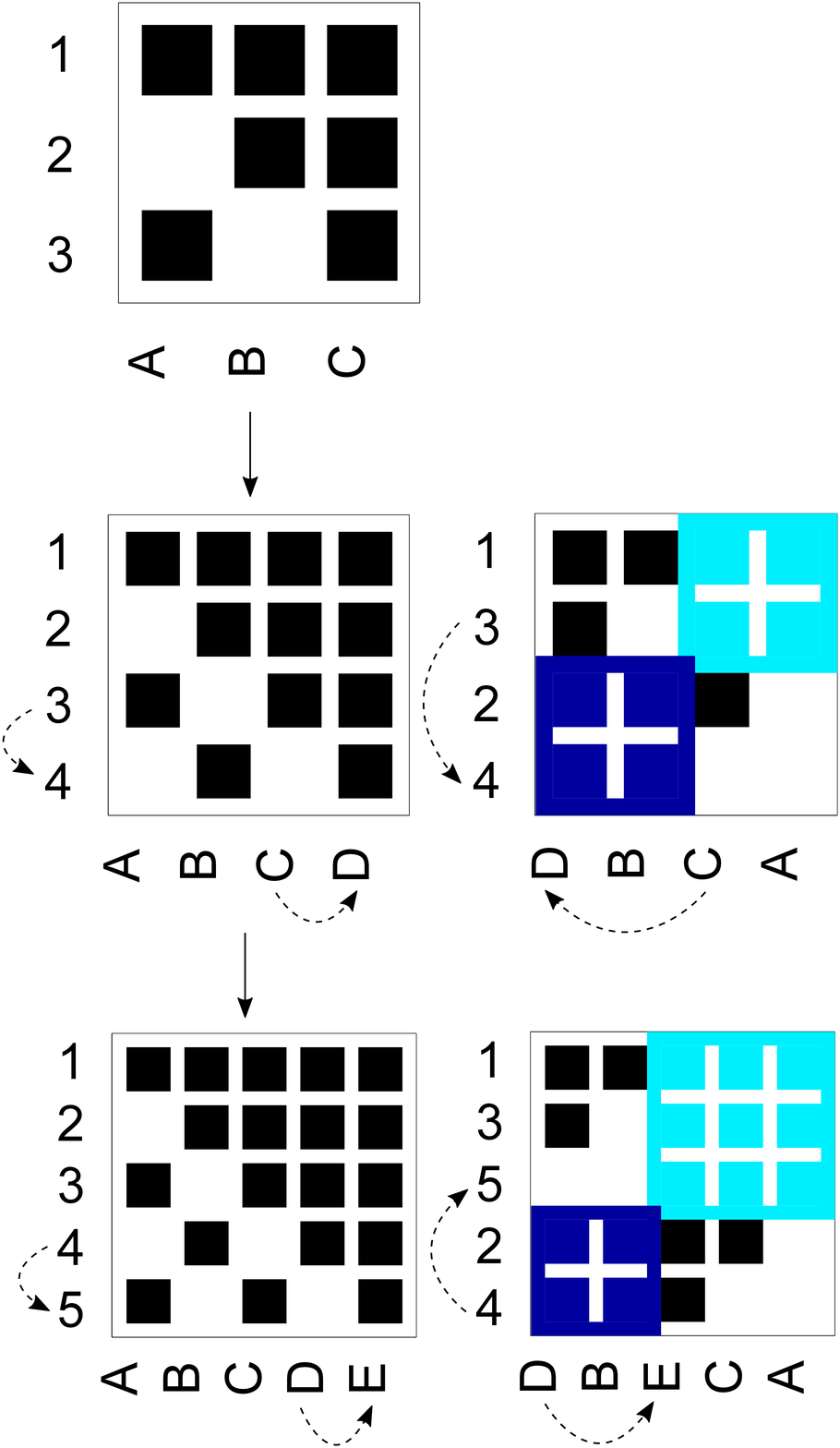}}
\caption{(Color online) Schematics of a realization of the network growth model for each asymptotic case. All networks started out as a single pair of interacting virus and host, and snapshots of the networks at $n=3$, $4$ and $5$ are shown. The parameters used are (a) $P_H=P_V=0$ (perfectly nested network), (b) $P_H=P_V=1$ (modular network), (c) $P_H=0$, $P_V=1$ (multi-scale nested-modular network), and (d) $P_H=1$, $P_V=0$ (superimposed nested-modular network). The solid arrows show the network evolution and dashed arrows denote duplications and go from ancestor to descendant. For (b) and (c) reordered network matrices are also shown to highlight the modularity.}
\label{fig4:net_sch}
\end{figure}

The initial steps of the network growth process for the four asymptotic cases are depicted in Fig. \ref{fig4:net_sch}. When $P_H=P_V=0$, there is no cost for the hosts and viruses to evolve new resistance and infectivity, respectively.  Hosts and viruses initiate an arms race as shown in Fig. \ref{fig4:net_sch}(a), wherein the newly evolved hosts and viruses are increasingly resistant and generalist respectively, and form a nested network structure. Figure \ref{fig4:net_sch}(b) shows the mechanisms of growth in the case of $P_H=P_V=1$. In this limit, the new host $H'$ is resistant to all viruses that exploited its ancestor but loses resistance to viruses previously unable to infect the ancestral host. As a consequence, the virus that can exploit $H'$  are necessarily not from the module associated with the ancestor.  Over the long-term, the newly evolved hosts and viruses switch between modules, and the growth process reinforces the perfectly modular structure with low nestedness. When $P_H=0$ and $P_V=1$ (Fig. \ref{fig4:net_sch}(c)), hosts develop resistance with no costs, but viruses exploit new host types at maximal costs.  This asymmetry implies that newly evolved virus switch between modules, while there is no such restriction for the host.  As such, hosts gain increasing resistance, leading to a nested pattern within each module. Finally, for $P_H=1$ and $P_V=0$ (Fig. \ref{fig4:net_sch}(d)) the opposite process takes place and newly mutated hosts switch modules while viruses continue to evolve a wider host range without bound. As a result, the hosts can only evade infection from a subset of viruses belonging to a different module. Each evolved host specializes in resistance against viruses in one of the modules with varying resistance range following a nested pattern. This process generates a fully-connected interaction pattern within each module and a nested pattern across different modules and yield the superimposed nested-modular structure. The networks generated in the asymptotic limits of the parameter space are informative with respect to the bulk of parameter space.  In Appendix A we use a perturbative approach to examine the expected level of nestedness in the limit $P_H\rightarrow 0$ and $P_V\rightarrow 0$ in addition to the expected level of modularity in the limit $P_H\rightarrow 1$ and $P_V\rightarrow 1$. We find that nestedness decreases smoothly from ${\mathcal{N}}_{NODF}=1$ and that modularity decreases smoothly from $Q=Q_{\mathrm{max}}$ near the limiting cases. We conclude that the generative mechanisms for nested and modular networks apply to both the asymptotic and bulk regimes. 

\subsection{Scaling of modularity and nestedness with network size}
\begin{figure}[!htbp]
\centering
%\vspace{5mm}
\subfigure[]{\includegraphics[width=5cm]{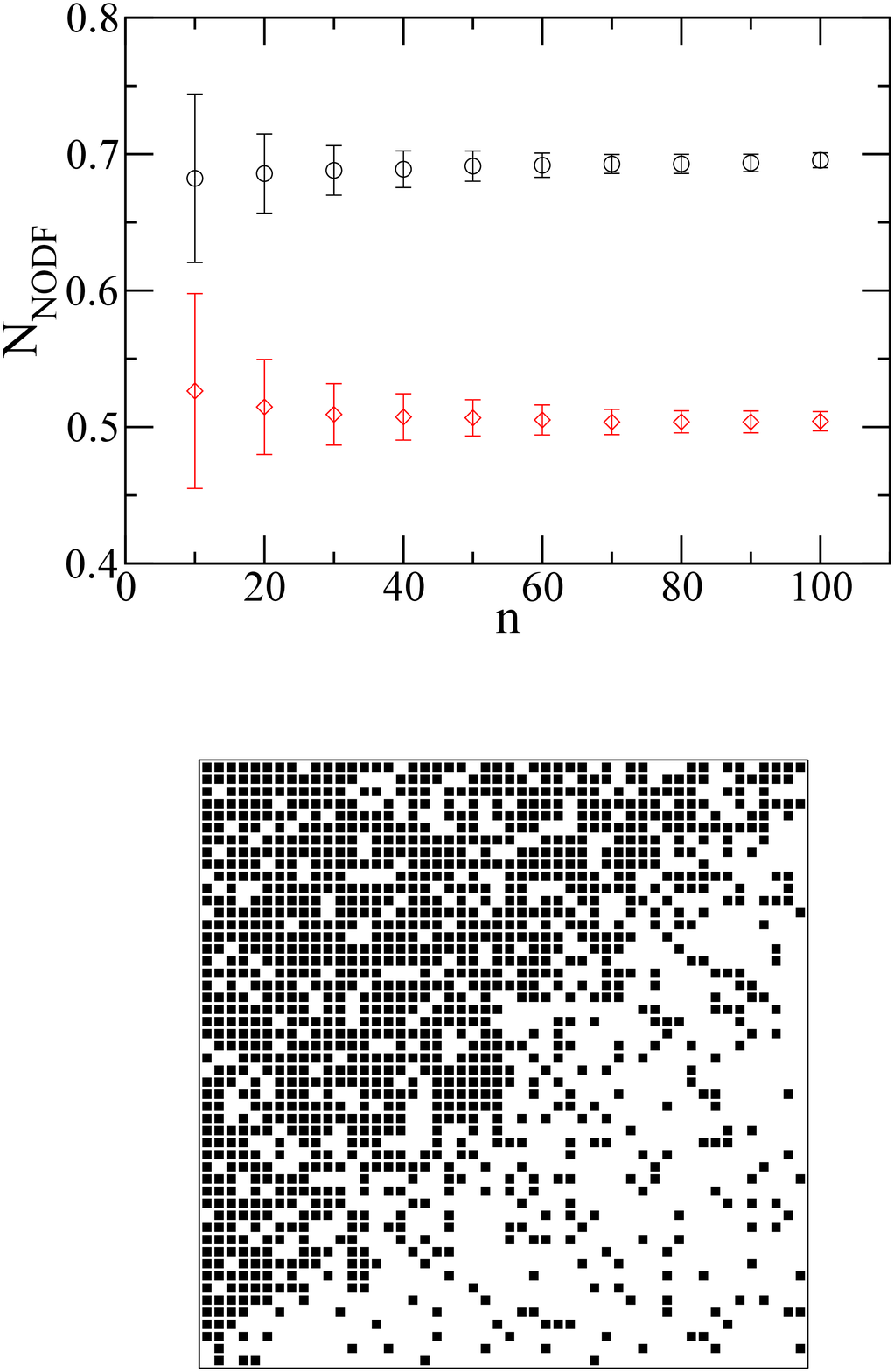}}
%\hspace{0.3cm}
\subfigure[]{\includegraphics[width=5cm]{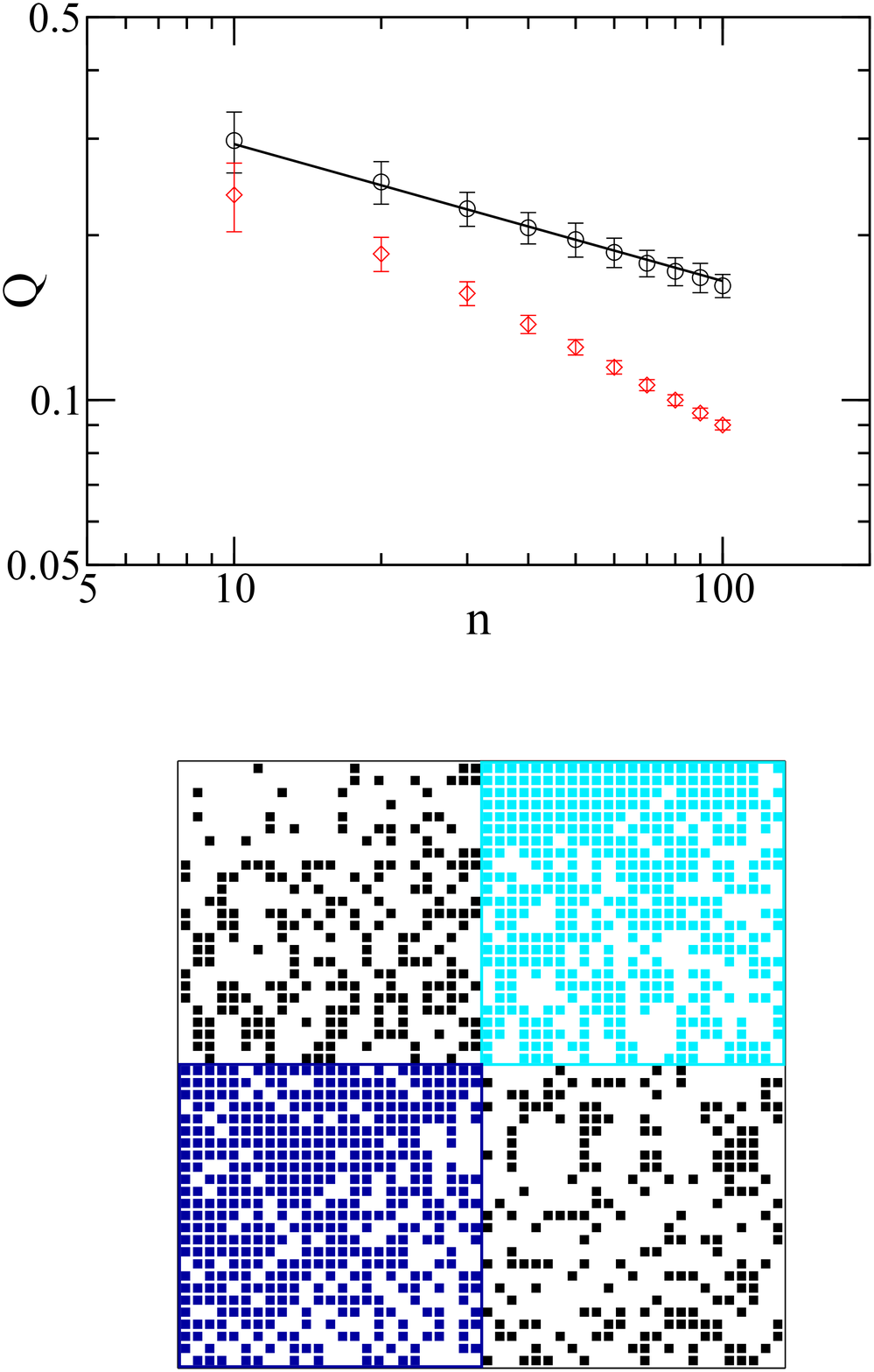}}
\hspace{3cm}
\subfigure[]{\includegraphics[width=5cm]{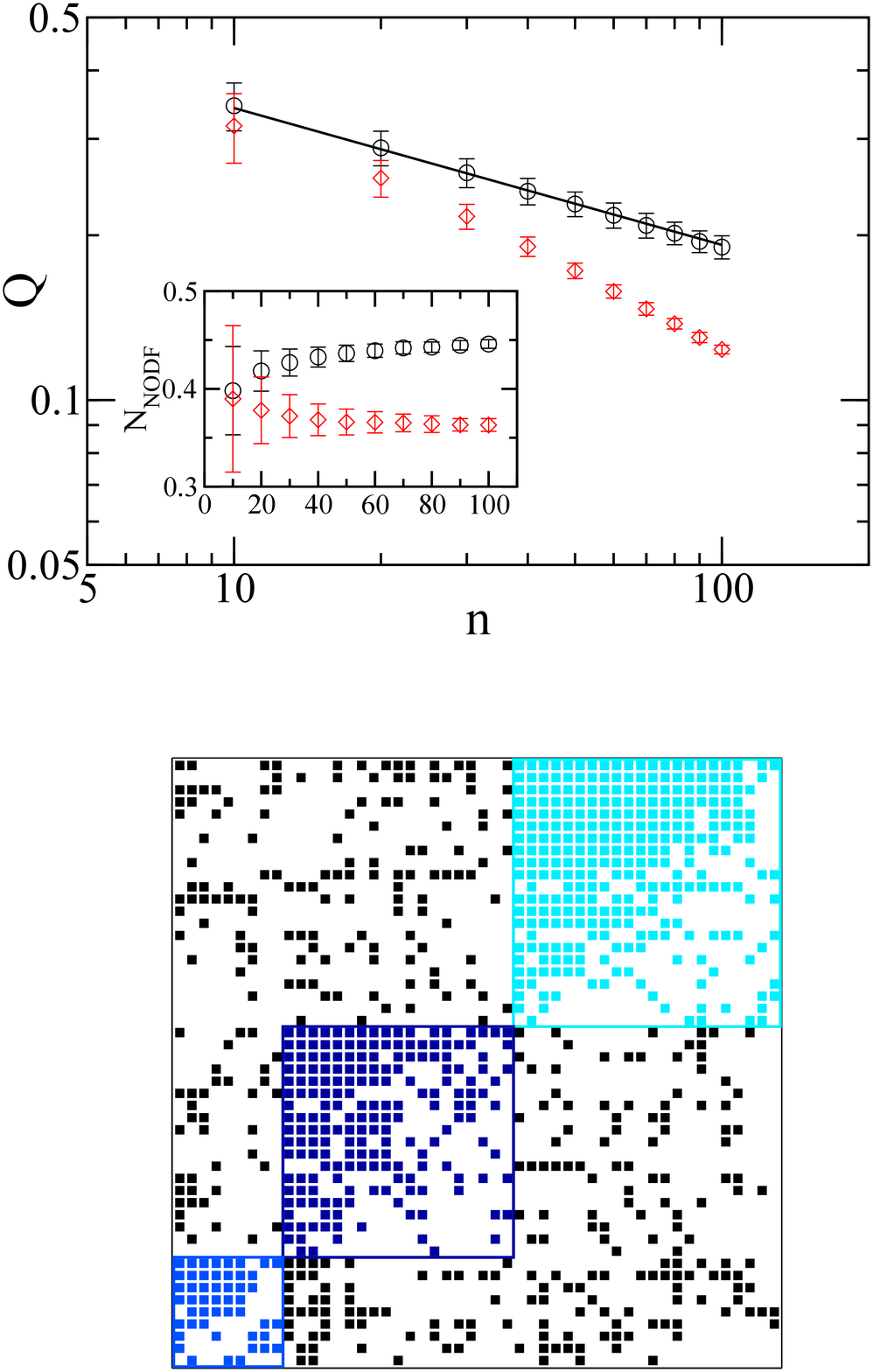}}
%\hspace{0.3cm}
\subfigure[]{\includegraphics[width=5cm]{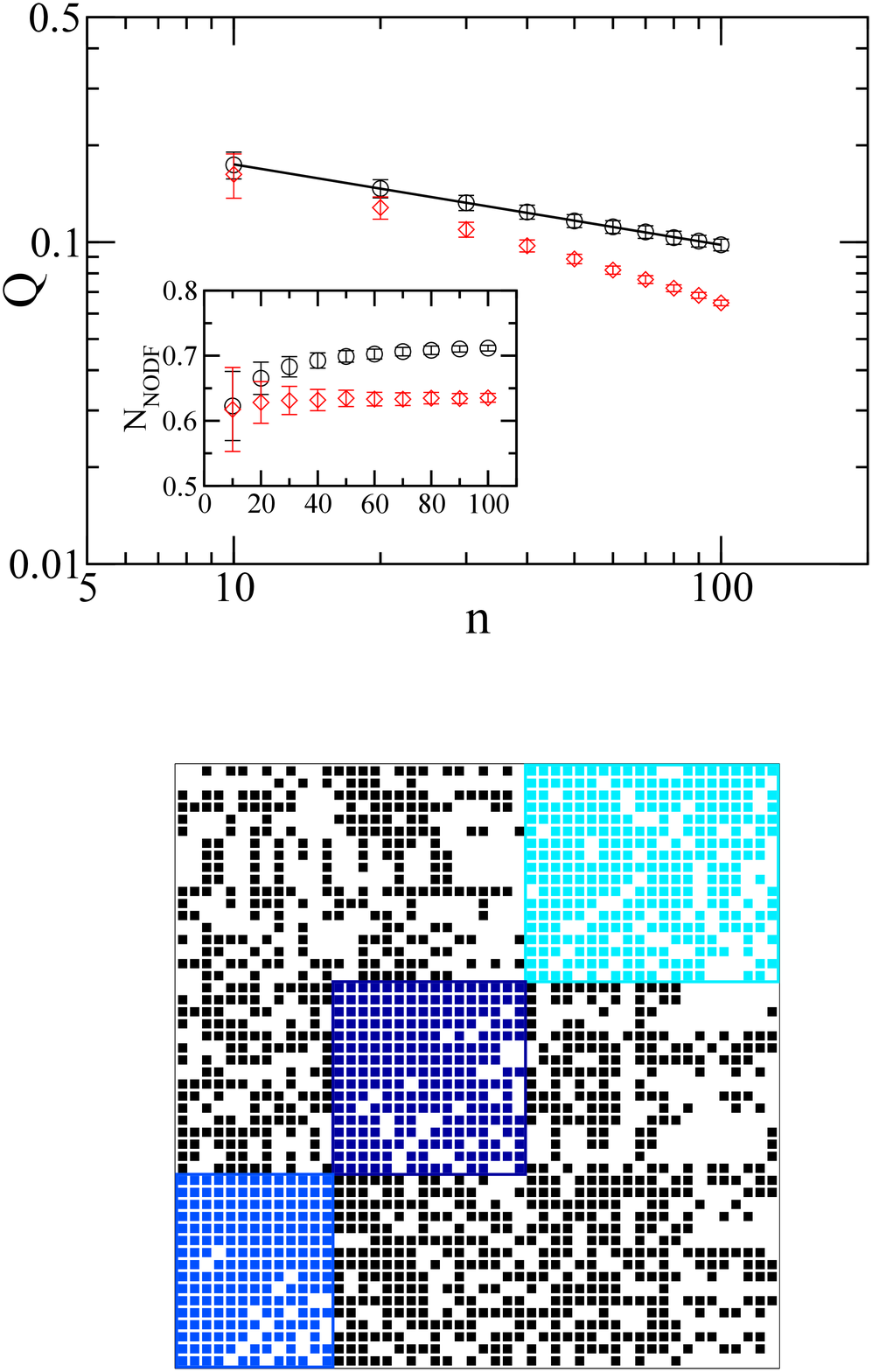}}
\caption{(Color online) Structural features of emergent networks as a function of network size. Network size dependence of (a) nestedness ${\mathcal{N}}_{NODF}$ of our model with $P_H=P_V=0.2$ (black circles) and Erd\H{o}s-R\'enyi (ER) random networks with the same connectance (red diamonds), and (b) modularity $Q$ of our model with $P_H=P_V=0.8$ (black circles) and the ER model (red diamonds). $Q$ (main panel) and ${\mathcal{N}}_{NODF}$ (inset) of our model with (c) $P_H=0.2$, $P_V=0.8$ and (d) $P_H=0.8$, $P_V=0.2$ (black circles) and their ER model counterparts (red diamonds). One realization of the network matrix for $n=50$ is shown for each set of parameters. Solid lines are power laws with the exponent $0.25$ and error bars have a width of $2$ standard deviations. Values of the modularity and nestedness for each data point are averaged over $10^6/n^2$ (rounded to the nearest integer) number of networks.}
\label{fig5:net_size}
\end{figure}

The bipartite network growth process continues unbounded.  As such, a full characterization of the emergent network structure requires an analysis of the scaling of modularity and nestedness as a function of network size. Fig. \ref{fig5:net_size} shows the size-dependence of modularity $Q$ and nestedness ${\mathcal{N}}_{NODF}$ for a subset of regions in parameter space. In Fig. \ref{fig5:net_size}(a) for $P_H=P_V=0.2$ near the perfectly nested asymptotic case, we see that our model gives a statistically significant nestedness throughout the range of network sizes compared to a null model of Erd\H{o}s-R\'enyi (ER) random network with the same connectance. This confirms that our model can generate and maintain significant nestedness when the network grows not only for the asymptotic case but also in the bulk of the parameter space. Figure \ref{fig5:net_size}(b) shows that our model gives statistically significant modularity for a set of parameters $P_H=P_V=0.8$ near the perfectly modular case when the network is sufficiently large. Although the modularity decreases with network size by a power law with exponent around $0.25$, the enhancement in modularity with respect to the null expectation increases for larger networks. For the parameter space region near the multi-scale nested-modular case, Fig. \ref{fig5:net_size}(c) presents the modularity and nestedness for the parameters $P_H=0.2$ and $P_V=0.8$. The results indicate elevated levels of both modularity and nestedness remain statistically significant after the network reaches a sufficiently large size. Figure \ref{fig5:net_size}(d) confirms that for $P_H=0.8$ and $P_V=0.2$ near the superimposed nested-modular case, the modularity and nestedness are also significant. However, there is a larger number of cross-module links than the multi-scale case as the interactions are not exclusively within modules even for the asymptotic case.

\subsection{Sensitivity of network structures to initial conditions}

\begin{figure}[!htbp]
\centering
\subfigure[]{\includegraphics[width=3.5cm]{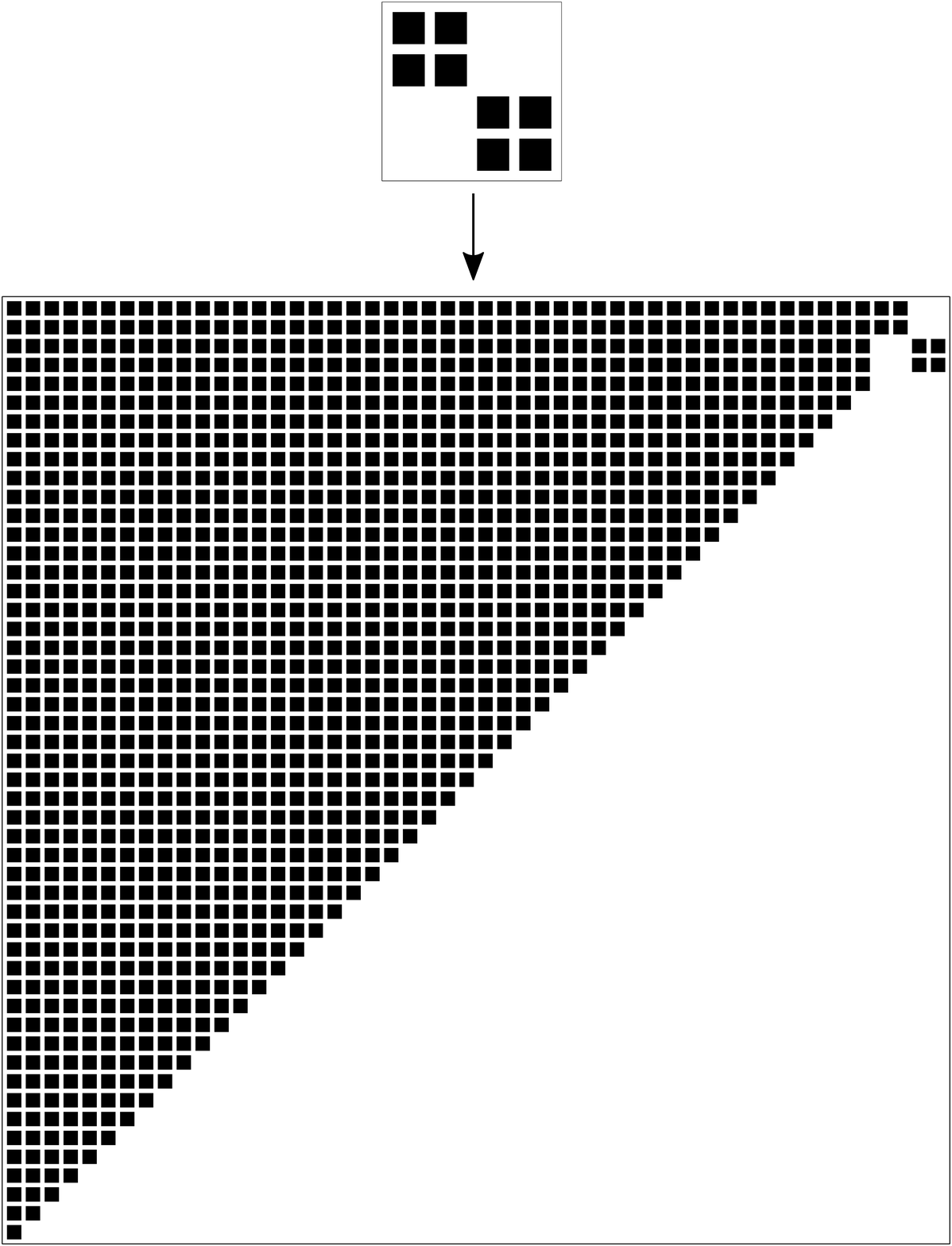}}
\hspace{0.3cm}
\subfigure[]{\includegraphics[width=3.5cm]{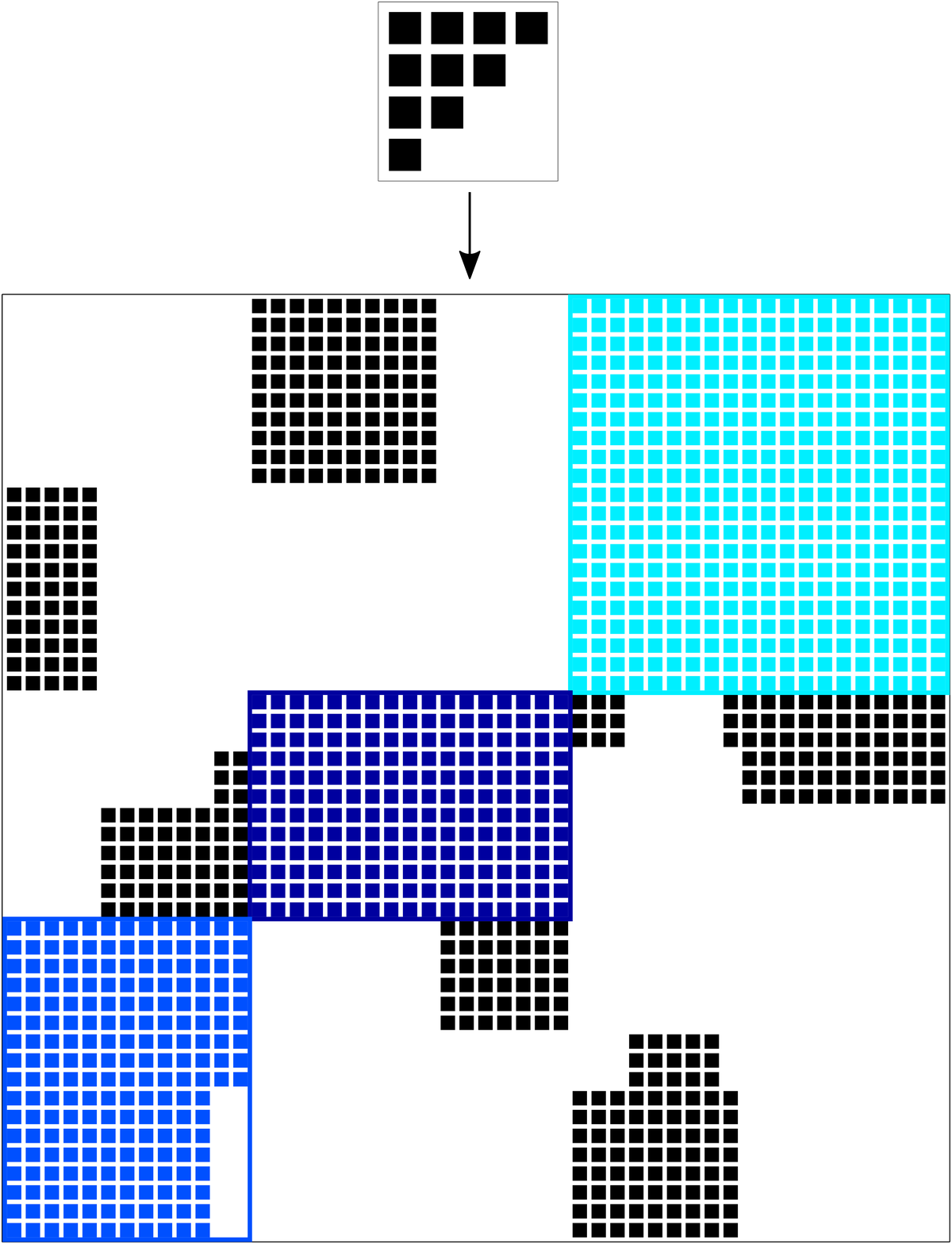}}
\hspace{0.3cm}
\subfigure[]{\includegraphics[width=3.5cm]{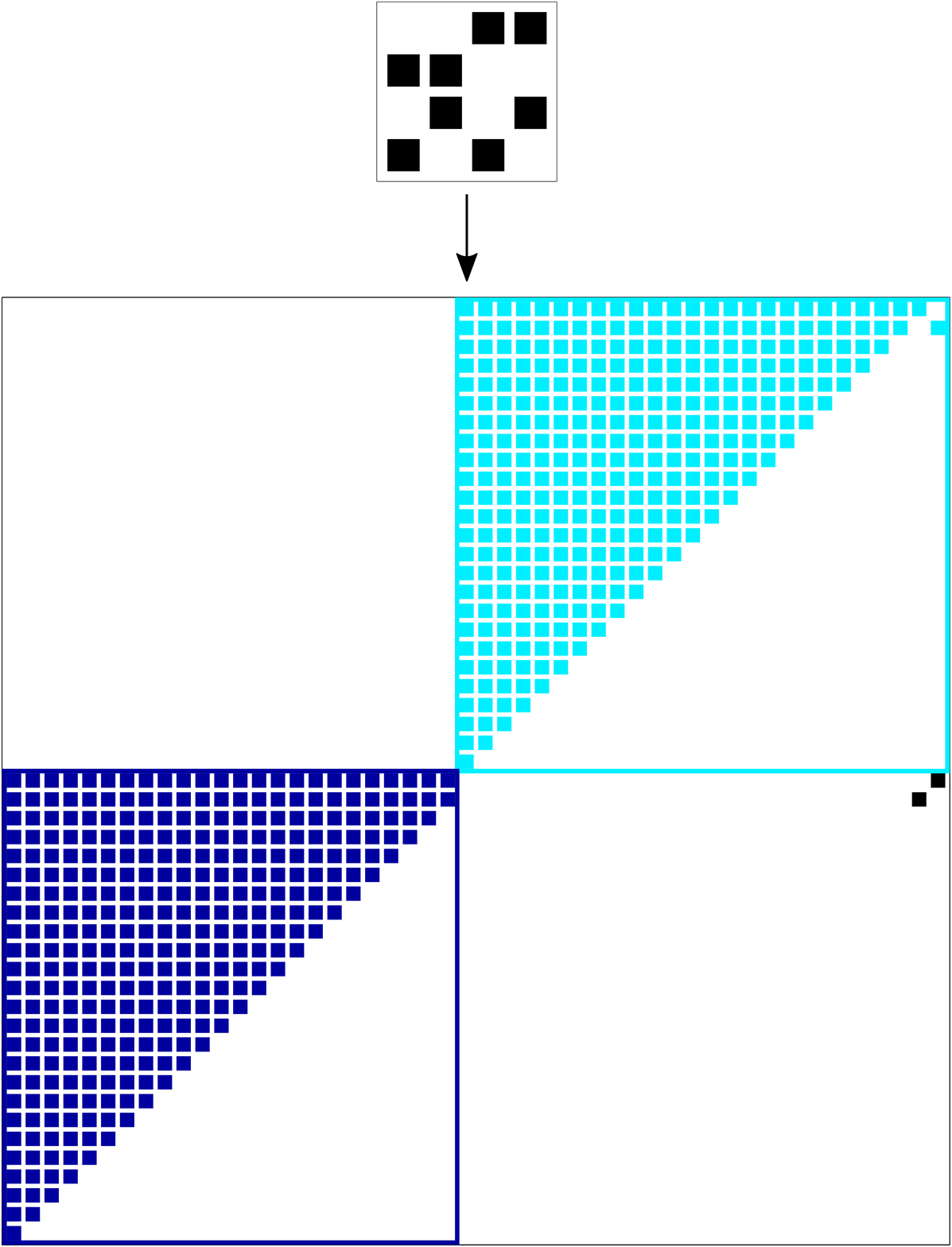}}
\hspace{0.3cm}
\subfigure[]{\includegraphics[width=3.5cm]{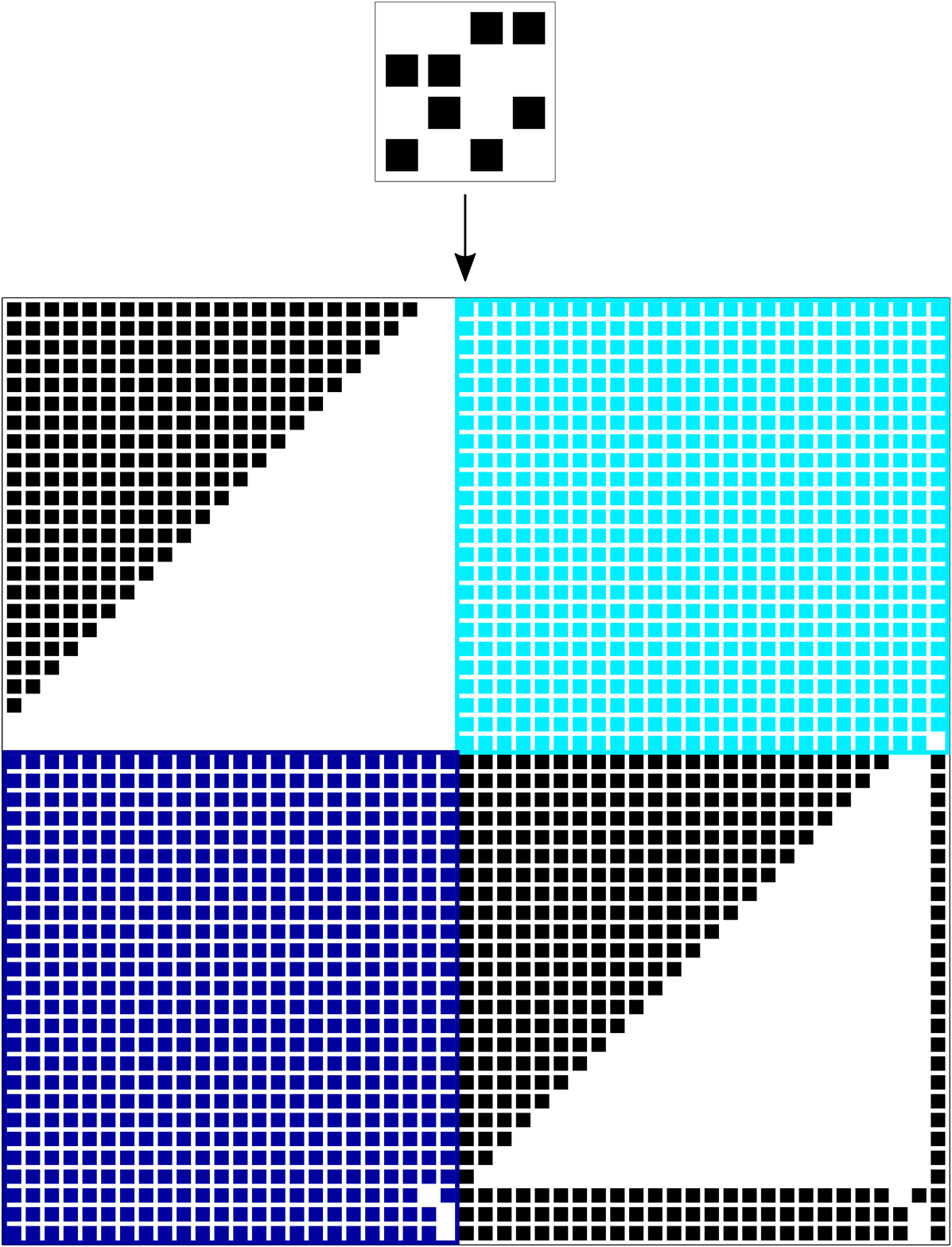}}
\caption{(Color online) Networks with asymptotic values of $P_H$ and $P_V$ generated from different initial network configurations: (a) $P_H=P_V=0$ (nested) generated from a perfectly modular initial network, (b) $P_H=P_V=1$ (modular) generated from a perfectly nested initial network, (c) $P_H=0$, $P_V=1$ (multi-scale nested-modular) from an initial network that minimizes the sum of modularity and nestedness at a fixed connectance of $0.5$, and (d) $P_H=1$, $P_V=0$ (superimposed nested-modular) from the same minimized network.}
\label{fig6:IC_net}
\end{figure}

We have so far focused on networks grown initially from a single pair of interacting host and virus strain as it is the typical way to initialize a coevolutionary experiment (see the review in Brockhurst \textit{et al.} \cite{brockhurst2013}). However, to investigate the robustness of the network structural patterns generated from our model, it is necessary to study also the sensitivity of these patterns to initial conditions. Figure \ref{fig6:IC_net} shows networks at the asymptotic cases of our model grown from alternative initial networks. In Fig. \ref{fig6:IC_net}(a), a perfectly modular initial network with $\mathcal{N}_{NODF}=0$ gives rise to a network with high nestedness ($\mathcal{N}_{NODF}=0.995$) for the nested asymptotic case ($P_H=P_V=0$). Similarly, Fig. \ref{fig6:IC_net}(b) shows that a perfectly nested initial network with low modularity ($Q=0.190$) yields a network with high modularity ($Q=0.314$) for the modular asymptotic case ($P_H=P_V=1$), albeit with more significant deviations from the perfectly modular network. In Figs. \ref{fig6:IC_net}(c) and (d) for the multi-scale and superimposed nested-modular case respectively, we choose an initial network that minimizes the sum of modularity and nestedness at a fixed connectance of $0.5$ obtained by a brute-force search. The minimized network ($\mathcal{N}_{NODF}=0$, $Q=0.313$) ensures that elevated levels of modularity and nestedness would be a result of the rules in generating the network instead of the initial conditions. In Figs. \ref{fig6:IC_net}(c) and (d), the multi-scale nested-modular ($P_H=0$, $P_V=1$) and superimposed nested-modular ($P_H=1$, $P_V=0$) networks are reproduced to good approximation with $\mathcal{N}_{NODF}=0.488$, $Q=0.497$ for the multi-scale case and $\mathcal{N}_{NODF}=0.844$, $Q=0.167$ for the superimposed case.

\begin{figure}[!htbp]
\centering
\subfigure[]{\includegraphics[width=8cm]{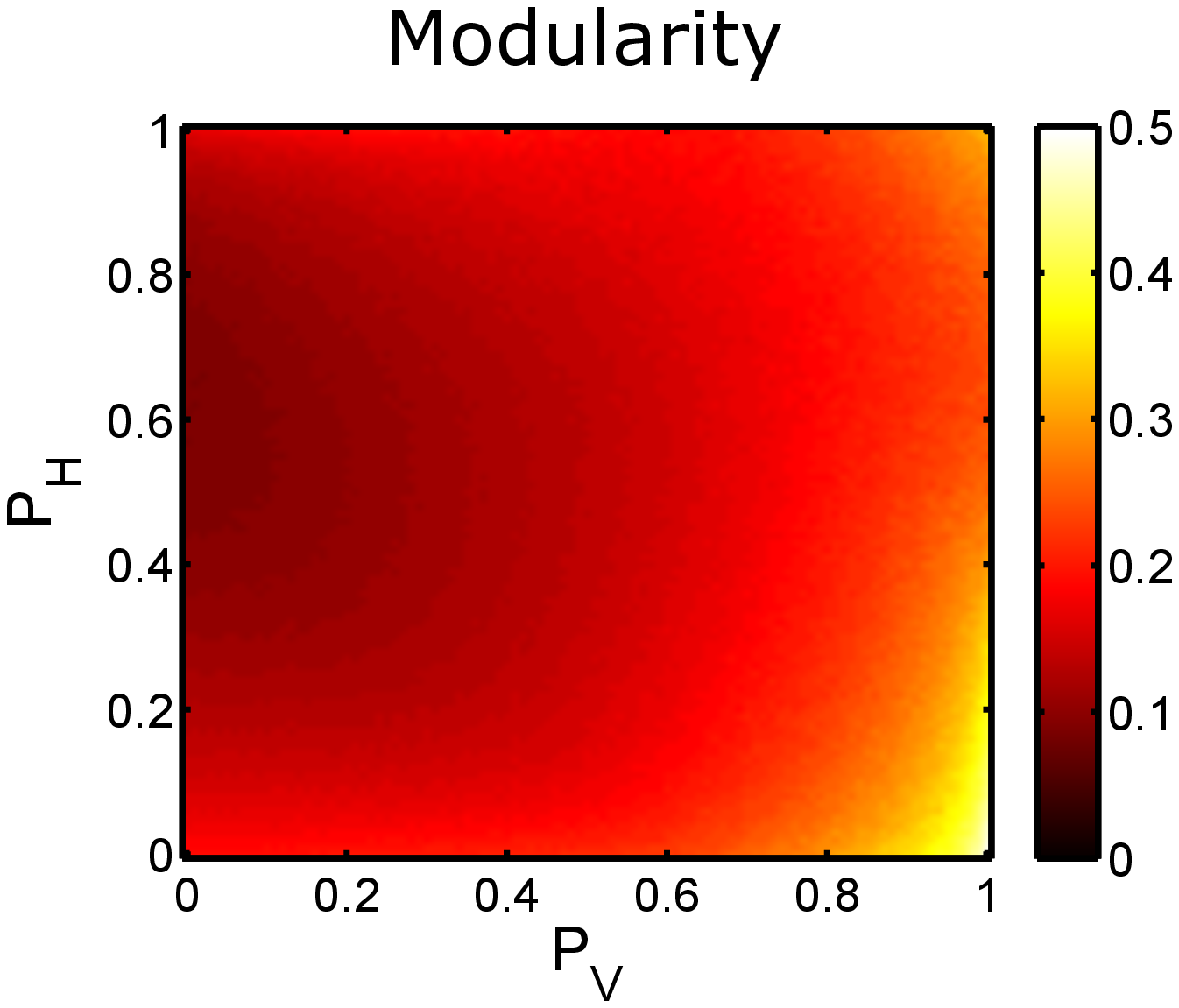}}
\subfigure[]{\includegraphics[width=8cm]{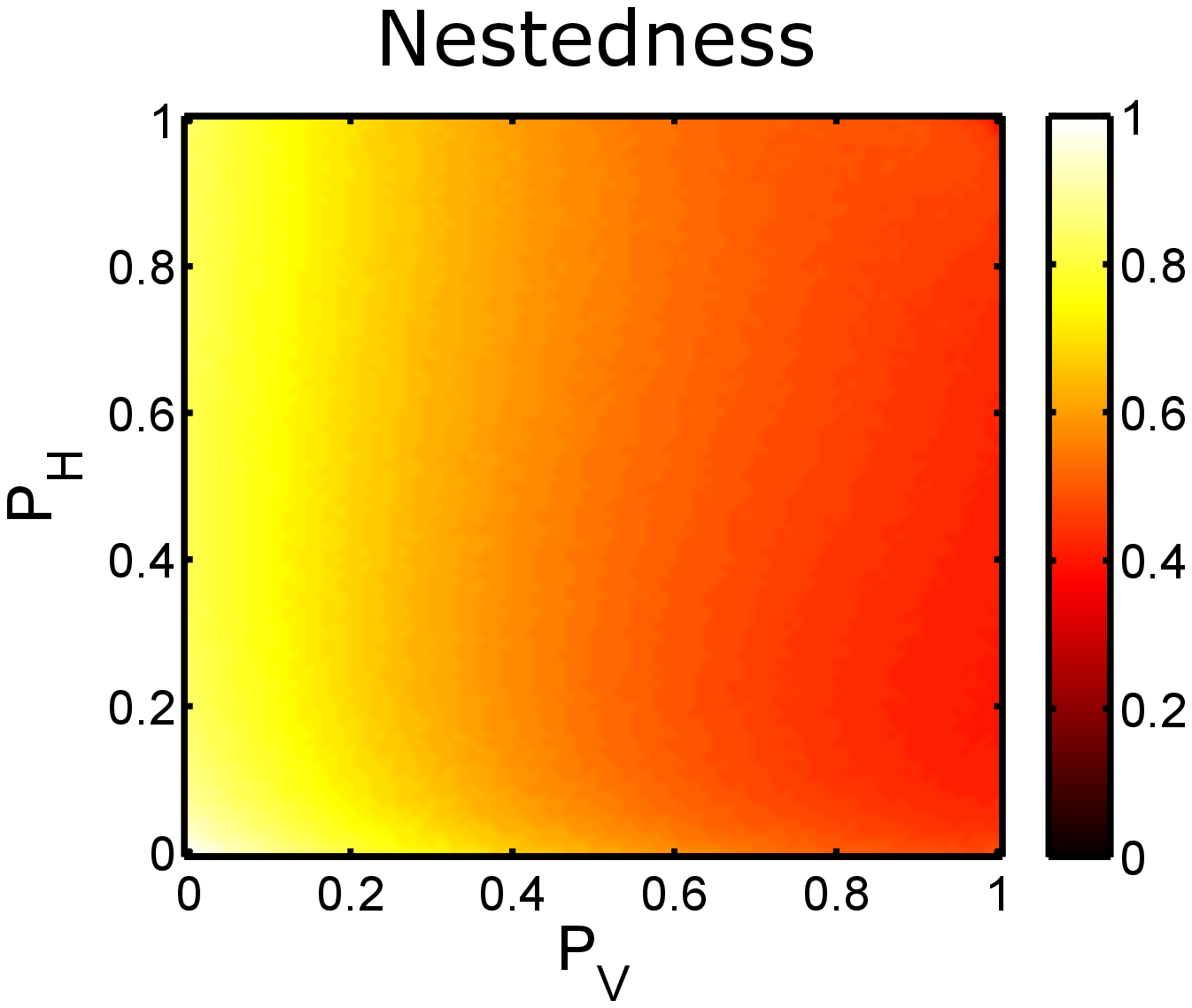}}
\subfigure[]{\includegraphics[width=8cm]{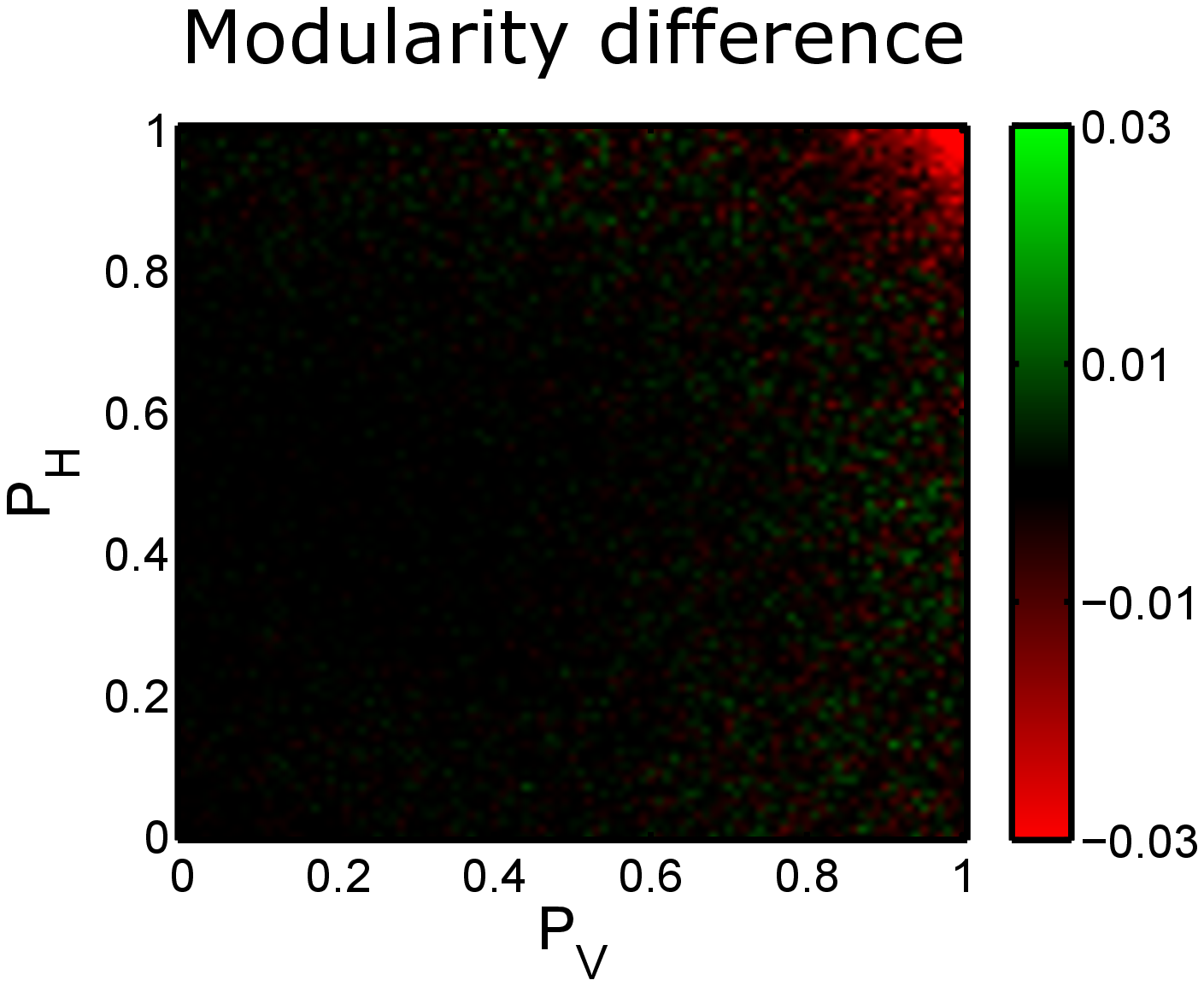}}
\subfigure[]{\includegraphics[width=8cm]{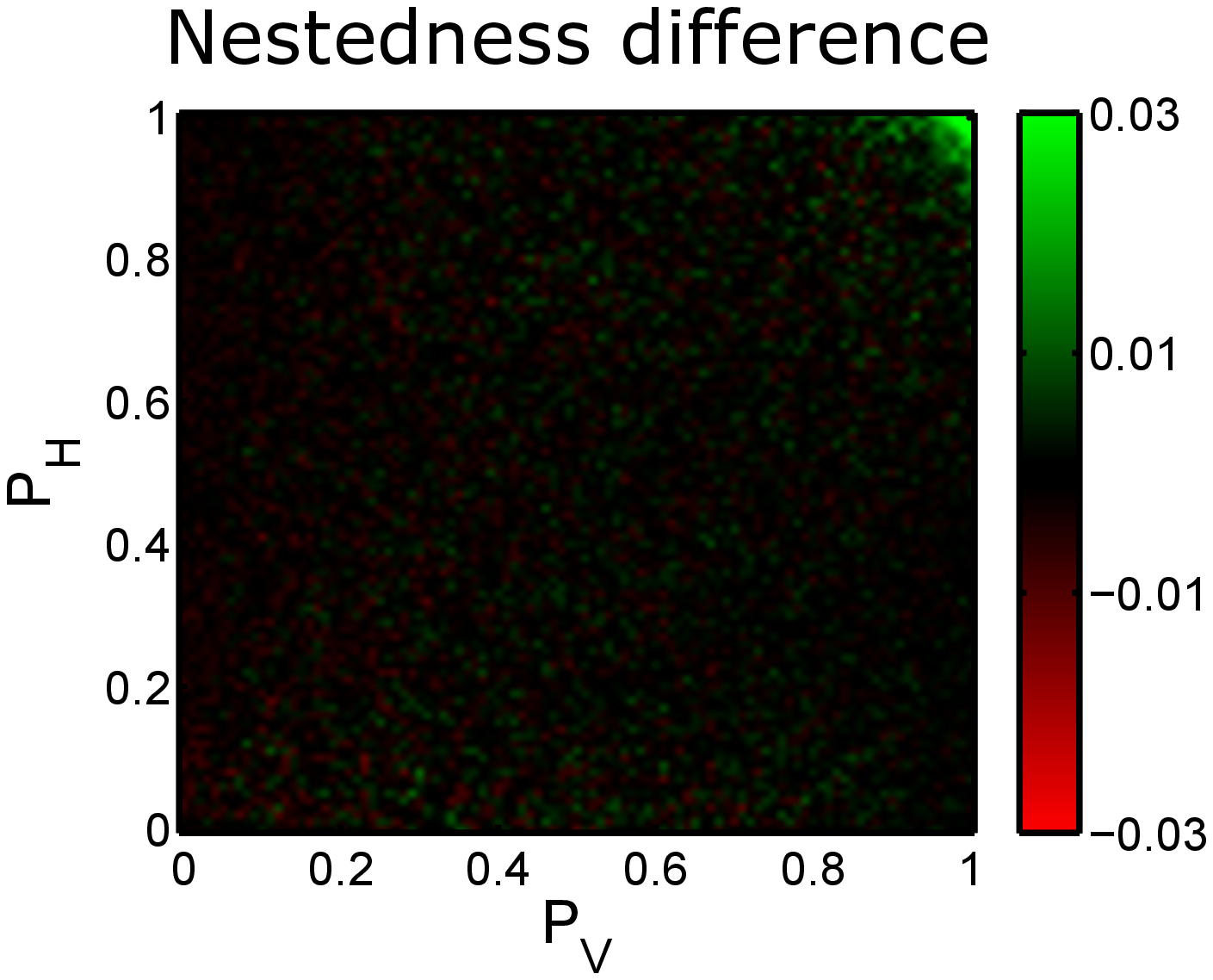}}
\caption{(Color online) Heat map showing the dependence of (a) modularity $Q$ and (b) nestedness ${\mathcal{N}}_{NODF}$ on the parameters $P_H$ and $P_V$ with initial networks generated from a modified form of ER random networks of average connectance $0.5$. The size of the initial networks is $4\times 4$ and that of the final networks is $50\times 50$. Values are averaged over $20$ independent realizations. Differences in (c) modularity and (d) nestedness between the random network initialization and the $1\times 1$ network initialization as used in Fig. \ref{fig2:phase}. Shades of green denote positive values (higher for random initialization) and shades of red indicate negative values.}
\label{fig7:IC_phase}
\end{figure}

To ascertain the robustness of network structural properties in the bulk of the parameter space, we reproduce the heat maps of modularity and nestedness with initial networks given by a modified form of ER random networks. We fix the diagonal elements of the network matrix to be $1$ and assign off-diagonal elements to be $0$ or $1$ randomly such that the overall average connectance is fixed. This is done to avoid unconnected nodes in the initial network as an unconnected virus would not be able to infect any host present. The results as shown in Figs. \ref{fig7:IC_phase} (a) and (b) confirm that the general trend of the dependence of modularity and nestedness remains unchanged. Figures \ref{fig7:IC_phase} (c) and (d) show the differences in modularity and nestedness between the random network initialization and the $1\times 1$ network initialization. The two types of initialization produce networks with very similar modularity and nestedness. Significant differences only appear near the modular asymptotic case of $P_H=P_V=1$ where random initialization leads to networks that are more nested and less modular, which is consistent with the observation in Fig. \ref{fig6:IC_net} that the non-nested modular network is less robust to change in initial conditions. The overall low sensitivity of the network patterns to initial conditions indicates that they arise as a result of the node selection and duplication rules independently of the initial conditions with the exception of a narrow region near the modular asymptotic case.

\section{Discussion}\label{sec:discuss}

We have developed a simple model of conflicting attachment for bipartite network growth inspired by virus-host interactions, thereby extending seminal models of preferential attachment~\cite{simon_1955,price_1976,barabasi1999}. The model shows different regimes of network structures including modular, nested, multi-scale nested-modular and a superimposed form of nested-modular structure. We show that these network structures are robust to network size and initial conditions. The mechanisms for the emergence of these structures can be understood by examining the costs to eliminating links and gaining links, by hosts and viruses respectively. When there are no costs for increasing resistance and infectivity, the hosts and viruses enter an arms race with increasingly resistant hosts and generalist viruses. This route to nestedness in an antagonistic system has parallels to the emergence of nestedness in mutualistic systems, e.g. plant-pollinator networks.  However, here nestedness emerges via conflict, i.e., the generalist virus infecting the most defended host, rather than mutualism, i.e., with both partners expanding their host ranges in tandem. When there are maximal costs for increasing resistance and range, newly added virus and hosts switch modules; this process reinforces the perfectly modular structure. When there are no costs for increasing resistance but maximal costs for developing new infectivity, the asymmetry leads to a case in which viruses switch modules while the hosts develop increasing resistance so that each module will have a maximally nested interaction structure. Lastly, when there are maximal costs for resistance and none for infectivity, the hosts switch modules while the viruses develop ever increasing host range, resulting in host strains that are infected by every virus within its module and develop varying levels of resistance to viruses from a different module.

Empirical studies of phage-bacteria systems have revealed nested and multi-scale structures which are reproduced in our model \cite{flores2011}. In contrast to preferential attachment mechanism that produces power-law degree distributions, our model produces uniform degree distributions for the asymptotic cases showing elevated nestedness as expected for highly nested networks. Network growth in these systems is a result of the coevolution between the phage and bacteria where new strains of phage and bacteria evolve under changing selective pressures due to the ever changing composition of phage and bacteria populations. Laboratory-based coevolution experiments have provided evidence that it is possible for phage to lose its ability to infect the ancestral host when gaining infectivity for contemporary bacterial strains \cite{scanlan2011}, as we have assumed in our model. In general, the comparative benefit of having a wide host range depends on the details of the trade-off between host range and replication rate, and it is possible for generalist and specialist viruses to coexist given certain conditions of trade-off \cite{jover_jtb2013}. However, it has been observed in coevolutionary experiments that phage strains tend to become more generalist over time \cite{scanlan2011}. It is therefore reasonable to assume that an increased host range confers an advantage in these systems, at least in the time scale ($\sim 20$ days) that has been studied. Nonetheless, generalizing the current approach to include other selection rules for network rewiring is essential.

In this manuscript we have focused on the statistical properties of an evolving network.  Despite the advantage of this approach being readily generalizable, it carries a distinct set of limitations and opportunities to incorporate increased realism.
First, due to the choice to model the dynamics of the network, the model cannot also quantify concurrent population dynamics.  The model also raises the question: to what extent can network connectivity be used as a proxy for relative fitness?  Prior ecological models show that when certain trade-offs are met, then having more links for hosts and fewer links for viruses may also correlate with ecological and even evolutionary success~\cite{jover_jtb2013,korytowski_2015}.  Yet, the evidence for the universality of such trade-offs, e.g., between resistance and growth rate, remains equivocal.
Another limitation is that nodes are continuously added to the network in the current conflicting attachment model - as is the case for the preferential attachment model. Such unbounded growth also has a direct biological analogue.  Coevolutionary experiments of viruses and microbes collect strains at different time points - the number of strains grows in the course of the experiment - and then assess the cross-infection of strains within and between time points~\cite{poullain_2008,scanlan2011,meyer_2012repeatability,sieber_isme2014}. However, the number of strains of organisms that can be supported at a given time in a given environment is constrained by ecological interactions and exogenous factors such as resource input rates.  Such constraints should be incorporated into future analyses of conflicting attachment if the emergent network is to be interpreted in terms of a dynamic, \emph{current state} of the system.

In moving forward, we consider it important to address both theoretical and empirical issues arising from this study. From a theoretical perspective, it will be important to consider changes to the current rewiring procedure such that probabilities of adding or removing links can depend on the degree of nodes. For example, in the kill-the-winner model of phage-bacteria population dynamics, newly evolved phages are more likely to infect a successful bacterial population that has a high enough population to support phage growth \cite{thingstad2000,winter2010}. Similarly, the selection criteria for duplication of virus and hosts can be extended to incorporate stochasticity and/or a degree-dependence on duplication. In a previously introduced model of virus-host coevolution \cite{rosvall2006}, strains of bacteria and phage are selected randomly with uniform probability for duplication. This random selection model and our conflicting attachment model are two limiting cases of a spectrum of selection mechanisms, in which speciation events are determined by pure randomness in the former case and by the node degree deterministically in the latter case. A more realistic selection scheme would fall within that spectrum where randomness plays a role but the node degree would influence the probability for a node to be duplicated. Furthermore, it is necessary to extend the model to weighted bipartite networks in order to study the effects of trade-offs, such as that between host range and infection rate studied in ecological models of phage-bacteria interactions. Different strategies of host exploitation and resistance can then be studied in the context of trade-offs by comparing agents with the same node strength (sum of weights of all links) but different node degrees. From an empirical perspective, the conflicting attachment model highlights the need to collect information on the time-course of changes in infectivity and resistance. Current design of coevolutionary experiments do not necessarily provide information on the community context under which hosts and viruses coevolve.   There are technical barriers to collecting such data, nonetheless, doing so would be an important step to addressing the question of the connection between the structure of infection networks and the fitness of coevolving populations.

\appendix*
\section{change in modularity and nestedness in the near-asymptotic cases}\label{sec:app}

In the main text, we demonstrated that the conflicting attachment model gives a perfectly modular network when $P_H=P_V=1$ and a perfectly nested network when $P_H=P_V=0$. Here we demonstrate that near the asymptotic cases, the modularity and nestedness changes continuously. This means that there is a crossover of the network properties instead of a discontinuous transition when the parameters move away from the asymptotic cases. This crossover behavior is illustrated for the two asymptotic cases with the following two examples.

Consider a perfectly modular network with two modules, $C_1$ and $C_2$, meaning that all hosts and viruses within a module are connected, but they do not interact across different modules. Such a configuration would maximize the modularity given the module sizes. For simplicity, further assume that each module has the same number of hosts and viruses given by $n_1$ and $n_2$ respectively for each module.

Now we apply the growth process of the conflicting attachment model to this network with the parameters $P_H=1-\epsilon$ and $P_V=1-\epsilon'$ where $\epsilon$ and $\epsilon'$ describe infinitesimal deviations from the asymptotic case. For $\epsilon =\epsilon'=0$, the network created after a network growth step would still be perfectly modular. The expected value of modularity after one network growth step is formally given by
\begin{equation}\label{eq:modu_exp}
\langle Q\rangle =\sum_i P_i(\epsilon,\epsilon')Q_i
\end{equation}
where $P_i(\epsilon,\epsilon')$ is the probability for the resulting network to have a modularity value of $Q_i$. Without loss of generality, let $Q_0$ be the maximum possible modularity of the resulting network, which corresponds to a perfectly modular network.

The perfectly modular structure can be broken either through the addition of a new host or a new virus in the network growth step. After a new host $H'$ is added by duplicating from an existing one in one of the modules, for example $C_1$, it severs all the links with viruses in $C_1$ and there is a probability $P_H=1-\epsilon$ for it to be linked to each virus in $C_2$. If $H'$ links to all viruses in $C_2$, the perfectly modular structure is preserved. Otherwise, the nodes in $C_2$ will no longer be fully connected as $H'$ is unconnected to some viruses in $C_2$. As $\epsilon$ is infinitesimally small, we consider that $H'$ is unconnected to at most one virus in $C_2$. The overall result is a network with a missing link in $C_2$ of the otherwise perfectly modular structure and has modularity $Q_1<Q_0$. The probability of this happening is $P_1(\epsilon) = n_2\epsilon(1-\epsilon)^{n_2-1}\approx n_2\epsilon$.

On the other hand, the addition of a new virus $V'$ has a different effect. Suppose $V'$ is duplicated from an existing virus in $C_1$, it will gain ability to exploit all the hosts in $C_2$. At the same time there is a chance of $\epsilon'$ that it will link to each host in $C_1$. Again using the perturbative argument, this results in a network with one cross-link between two modules and a modularity value of $Q_2<Q_0$ occurring with probability $P_2(\epsilon')=n_1\epsilon'(1-\epsilon')^{n_1-1}\approx n_1\epsilon'$. We further approximate the probability for the perfectly modular structure to be preserved as $P_0(\epsilon,\epsilon')\approx (1-P_1-P_2)\approx 1-n_2\epsilon-n_1\epsilon'$. Therefore from Eq. (\ref{eq:modu_exp}), we have
\begin{eqnarray}\label{eq:modu_exp_simp}
\langle Q\rangle&=&[1-P_1(\epsilon)-P_2(\epsilon')]Q_0+P_1(\epsilon)Q_1+P_2(\epsilon')Q_2
\nonumber \\
&\approx& Q_0-n_2\epsilon(Q_0-Q_1)-n_1\epsilon'(Q_0-Q_2)
\end{eqnarray}
Since $Q_0>Q_1$ and $Q_2$, the expected value of modularity after one network growth step is
decreased from the perfectly modular value of $Q_0$ and the differences are proportional to $\epsilon$ and $\epsilon'$. In particular, when $\epsilon$ and $\epsilon'$ both approach zero, $\langle Q\rangle$ increases continuously and approaches $Q_0$.

Next we consider another near-asymptotic case with $P_H=\epsilon$ and $P_V=\epsilon'$. It reduces to the perfectly nested case when $\epsilon=\epsilon'=0$. Starting with an $n\times n$ perfectly nested network, again we consider the effect of a single network growth step. Similar to Eq. (\ref{eq:modu_exp}), the expected value of the nestedness after the growth is given by
\begin{equation}\label{eq:nest_exp}
\langle {\mathcal{N}}\rangle =\sum_i {\mathcal{P}}_i(\epsilon,\epsilon')\mathcal{N}_i
\end{equation}
where ${\mathcal{P}}_i(\epsilon,\epsilon')$ is the probability of the resultant network having a nestedness of ${\mathcal{N}}_i$. 

First consider the effect of the addition of a host to the network structure. The selection criteria for duplication of the model means that the most resistant host $H_r$ (lowest degree) would be chosen for duplication. Given that the network is nested, a newly added host $H'$ would develop resistance to the most generalist virus $V_g$ (highest degree) but have a probability $\epsilon$ to reconnect to any other existing virus. Again assuming $\epsilon$ to be infinitesimally small, the process either preserves the perfectly nested structure with ${\mathcal{N}}_0=1$ or creates an extra link between $H'$ and a virus $V_q$ ($q\neq g$) in an otherwise perfectly nested structure with a probability of $\epsilon$. The nestedness of the latter structure depends on $q$ and is denoted by ${\mathcal{N}}_1(q)<1$.

The addition of a new virus $V'$ is a duplication of the most generalist virus $V_g$. $V'$ then connects to $H'$ and has a probability $1-\epsilon'$ to continue to exploit any other host. Due to the small value of $\epsilon'$, this results in either the nested structure being preserved or a single missing link between $V'$ and a host $H_{q\,'}$ ($q\,'\neq N$) with probability $\epsilon'$ and a nestedness of ${\mathcal{N}}_2(q\,')<1$. Thus we obtain an estimate for the expected value of nestedness
\begin{eqnarray}\label{eq:nest_exp_simp}
\langle {\mathcal{N}}\rangle &\approx& [1-(n-1)\epsilon-(n-1)\epsilon']{\mathcal{N}}_0
+\epsilon\sum_{q\neq g}{\mathcal{N}}_1(q)+\epsilon' \sum_{q\,'\neq N}{\mathcal{N}}_2(q\,')
\nonumber \\
&=& {\mathcal{N}}_0-\epsilon\sum_{q\neq g}[{\mathcal{N}}_0-{\mathcal{N}}_1(q)]-\epsilon'\sum_{q\,'\neq N}[{\mathcal{N}}_0-{\mathcal{N}}_2(q\,')].
\end{eqnarray}
As the perfectly nested network has the maximum possible nestedness value of ${\mathcal{N}}_0=1$, ${\mathcal{N}}_0-{\mathcal{N}}_1(q)>0$ and ${\mathcal{N}}_0-{\mathcal{N}}_2(q\,')>0$. Hence, the expected value of nestedness decreases continuously as one deviates from the asymptotic case.

\begin{acknowledgments}
This work is supported by Army Research Office grant W911NF-14-1-0402.
The authors thank Justin Meyer for input on the manuscript.
\end{acknowledgments}

\bibliography{reference}{}

%merlin.mbs apsrev4-1.bst 2010-07-25 4.21a (PWD, AO, DPC) hacked
%Control: key (0)
%Control: author (72) initials jnrlst
%Control: editor formatted (1) identically to author
%Control: production of article title (-1) disabled
%Control: page (0) single
%Control: year (1) truncated
%Control: production of eprint (0) enabled
\begin{thebibliography}{50}%
\makeatletter
\providecommand \@ifxundefined [1]{%
 \@ifx{#1\undefined}
}%
\providecommand \@ifnum [1]{%
 \ifnum #1\expandafter \@firstoftwo
 \else \expandafter \@secondoftwo
 \fi
}%
\providecommand \@ifx [1]{%
 \ifx #1\expandafter \@firstoftwo
 \else \expandafter \@secondoftwo
 \fi
}%
\providecommand \natexlab [1]{#1}%
\providecommand \enquote  [1]{``#1''}%
\providecommand \bibnamefont  [1]{#1}%
\providecommand \bibfnamefont [1]{#1}%
\providecommand \citenamefont [1]{#1}%
\providecommand \href@noop [0]{\@secondoftwo}%
\providecommand \href [0]{\begingroup \@sanitize@url \@href}%
\providecommand \@href[1]{\@@startlink{#1}\@@href}%
\providecommand \@@href[1]{\endgroup#1\@@endlink}%
\providecommand \@sanitize@url [0]{\catcode `\\12\catcode `\$12\catcode
  `\&12\catcode `\#12\catcode `\^12\catcode `\_12\catcode `\%12\relax}%
\providecommand \@@startlink[1]{}%
\providecommand \@@endlink[0]{}%
\providecommand \url  [0]{\begingroup\@sanitize@url \@url }%
\providecommand \@url [1]{\endgroup\@href {#1}{\urlprefix }}%
\providecommand \urlprefix  [0]{URL }%
\providecommand \Eprint [0]{\href }%
\providecommand \doibase [0]{http://dx.doi.org/}%
\providecommand \selectlanguage [0]{\@gobble}%
\providecommand \bibinfo  [0]{\@secondoftwo}%
\providecommand \bibfield  [0]{\@secondoftwo}%
\providecommand \translation [1]{[#1]}%
\providecommand \BibitemOpen [0]{}%
\providecommand \bibitemStop [0]{}%
\providecommand \bibitemNoStop [0]{.\EOS\space}%
\providecommand \EOS [0]{\spacefactor3000\relax}%
\providecommand \BibitemShut  [1]{\csname bibitem#1\endcsname}%
\let\auto@bib@innerbib\@empty
%</preamble>
\bibitem [{\citenamefont {Simon}(1955)}]{simon_1955}%
  \BibitemOpen
  \bibfield  {author} {\bibinfo {author} {\bibfnamefont {H.~A.}\ \bibnamefont
  {Simon}},\ }\href {\doibase 10.1093/biomet/42.3-4.425} {\bibfield  {journal}
  {\bibinfo  {journal} {Biometrika}\ }\textbf {\bibinfo {volume} {42}},\
  \bibinfo {pages} {425} (\bibinfo {year} {1955})}\BibitemShut {NoStop}%
\bibitem [{\citenamefont {Price}(1976)}]{price_1976}%
  \BibitemOpen
  \bibfield  {author} {\bibinfo {author} {\bibfnamefont {D.~D.~S.}\
  \bibnamefont {Price}},\ }\href {\doibase 10.1002/asi.4630270505} {\bibfield
  {journal} {\bibinfo  {journal} {J. Assoc. Inf. Sci. Technol.}\ }\textbf
  {\bibinfo {volume} {27}},\ \bibinfo {pages} {292} (\bibinfo {year}
  {1976})}\BibitemShut {NoStop}%
\bibitem [{\citenamefont {Barab{\'a}si}\ and\ \citenamefont
  {Albert}(1999)}]{barabasi1999}%
  \BibitemOpen
  \bibfield  {author} {\bibinfo {author} {\bibfnamefont {A.-L.}\ \bibnamefont
  {Barab{\'a}si}}\ and\ \bibinfo {author} {\bibfnamefont {R.}~\bibnamefont
  {Albert}},\ }\href {\doibase 10.1126/science.286.5439.509} {\bibfield
  {journal} {\bibinfo  {journal} {Science}\ }\textbf {\bibinfo {volume}
  {286}},\ \bibinfo {pages} {509} (\bibinfo {year} {1999})}\BibitemShut
  {NoStop}%
\bibitem [{\citenamefont {Albert}\ and\ \citenamefont
  {Barab{\'a}si}(2002)}]{albert2002}%
  \BibitemOpen
  \bibfield  {author} {\bibinfo {author} {\bibfnamefont {R.}~\bibnamefont
  {Albert}}\ and\ \bibinfo {author} {\bibfnamefont {A.-L.}\ \bibnamefont
  {Barab{\'a}si}},\ }\href {\doibase 10.1103/RevModPhys.74.47} {\bibfield
  {journal} {\bibinfo  {journal} {Rev. Mod. Phys.}\ }\textbf {\bibinfo {volume}
  {74}},\ \bibinfo {pages} {47} (\bibinfo {year} {2002})}\BibitemShut {NoStop}%
\bibitem [{\citenamefont {Barab{\'a}si}\ \emph {et~al.}(2000)\citenamefont
  {Barab{\'a}si}, \citenamefont {Albert},\ and\ \citenamefont
  {Jeong}}]{barabasi2000}%
  \BibitemOpen
  \bibfield  {author} {\bibinfo {author} {\bibfnamefont {A.-L.}\ \bibnamefont
  {Barab{\'a}si}}, \bibinfo {author} {\bibfnamefont {R.}~\bibnamefont
  {Albert}}, \ and\ \bibinfo {author} {\bibfnamefont {H.}~\bibnamefont
  {Jeong}},\ }\href {\doibase 10.1016/S0378-4371(00)00018-2} {\bibfield
  {journal} {\bibinfo  {journal} {Phys. A}\ }\textbf {\bibinfo {volume}
  {281}},\ \bibinfo {pages} {69} (\bibinfo {year} {2000})}\BibitemShut
  {NoStop}%
\bibitem [{\citenamefont {Eguiluz}\ \emph {et~al.}(2005)\citenamefont
  {Eguiluz}, \citenamefont {Chialvo}, \citenamefont {Cecchi}, \citenamefont
  {Baliki},\ and\ \citenamefont {Apkarian}}]{eguiluz2005}%
  \BibitemOpen
  \bibfield  {author} {\bibinfo {author} {\bibfnamefont {V.~M.}\ \bibnamefont
  {Eguiluz}}, \bibinfo {author} {\bibfnamefont {D.~R.}\ \bibnamefont
  {Chialvo}}, \bibinfo {author} {\bibfnamefont {G.~A.}\ \bibnamefont {Cecchi}},
  \bibinfo {author} {\bibfnamefont {M.}~\bibnamefont {Baliki}}, \ and\ \bibinfo
  {author} {\bibfnamefont {A.~V.}\ \bibnamefont {Apkarian}},\ }\href {\doibase
  10.1103/PhysRevLett.94.018102} {\bibfield  {journal} {\bibinfo  {journal}
  {Phys. Rev. Lett.}\ }\textbf {\bibinfo {volume} {94}},\ \bibinfo {pages}
  {018102} (\bibinfo {year} {2005})}\BibitemShut {NoStop}%
\bibitem [{\citenamefont {Abe}\ and\ \citenamefont {Suzuki}(2004)}]{abe2004}%
  \BibitemOpen
  \bibfield  {author} {\bibinfo {author} {\bibfnamefont {S.}~\bibnamefont
  {Abe}}\ and\ \bibinfo {author} {\bibfnamefont {N.}~\bibnamefont {Suzuki}},\
  }\href {\doibase 10.1209/epl/i2003-10108-1} {\bibfield  {journal} {\bibinfo
  {journal} {EPL (Europhys. Lett.)}\ }\textbf {\bibinfo {volume} {65}},\
  \bibinfo {pages} {581} (\bibinfo {year} {2004})}\BibitemShut {NoStop}%
\bibitem [{\citenamefont {Newman}(2001{\natexlab{a}})}]{newman2001}%
  \BibitemOpen
  \bibfield  {author} {\bibinfo {author} {\bibfnamefont {M.~E.~J.}\
  \bibnamefont {Newman}},\ }\href {\doibase 10.1103/PhysRevE.64.025102}
  {\bibfield  {journal} {\bibinfo  {journal} {Phys. Rev. E}\ }\textbf {\bibinfo
  {volume} {64}},\ \bibinfo {pages} {025102} (\bibinfo {year}
  {2001}{\natexlab{a}})}\BibitemShut {NoStop}%
\bibitem [{\citenamefont {Jeong}\ \emph {et~al.}(2003)\citenamefont {Jeong},
  \citenamefont {N{\'e}da},\ and\ \citenamefont {Barab{\'a}si}}]{jeong2003}%
  \BibitemOpen
  \bibfield  {author} {\bibinfo {author} {\bibfnamefont {H.}~\bibnamefont
  {Jeong}}, \bibinfo {author} {\bibfnamefont {Z.}~\bibnamefont {N{\'e}da}}, \
  and\ \bibinfo {author} {\bibfnamefont {A.-L.}\ \bibnamefont {Barab{\'a}si}},\
  }\href {\doibase 10.1209/epl/i2003-00166-9} {\bibfield  {journal} {\bibinfo
  {journal} {EPL (Europhys. Lett.)}\ }\textbf {\bibinfo {volume} {61}},\
  \bibinfo {pages} {567} (\bibinfo {year} {2003})}\BibitemShut {NoStop}%
\bibitem [{\citenamefont {Eisenberg}\ and\ \citenamefont
  {Levanon}(2003)}]{eisenberg2003}%
  \BibitemOpen
  \bibfield  {author} {\bibinfo {author} {\bibfnamefont {E.}~\bibnamefont
  {Eisenberg}}\ and\ \bibinfo {author} {\bibfnamefont {E.~Y.}\ \bibnamefont
  {Levanon}},\ }\href {\doibase 10.1103/PhysRevLett.91.138701} {\bibfield
  {journal} {\bibinfo  {journal} {Phys. Rev. Lett.}\ }\textbf {\bibinfo
  {volume} {91}},\ \bibinfo {pages} {138701} (\bibinfo {year}
  {2003})}\BibitemShut {NoStop}%
\bibitem [{\citenamefont {Ren}\ \emph {et~al.}(2012)\citenamefont {Ren},
  \citenamefont {Shen},\ and\ \citenamefont {Cheng}}]{ren2012}%
  \BibitemOpen
  \bibfield  {author} {\bibinfo {author} {\bibfnamefont {F.-X.}\ \bibnamefont
  {Ren}}, \bibinfo {author} {\bibfnamefont {H.-W.}\ \bibnamefont {Shen}}, \
  and\ \bibinfo {author} {\bibfnamefont {X.-Q.}\ \bibnamefont {Cheng}},\ }\href
  {\doibase 10.1016/j.physa.2012.02.001} {\bibfield  {journal} {\bibinfo
  {journal} {Phys. A}\ }\textbf {\bibinfo {volume} {391}},\ \bibinfo {pages}
  {3533} (\bibinfo {year} {2012})}\BibitemShut {NoStop}%
\bibitem [{\citenamefont {H{\'e}bert-Dufresne}\ \emph
  {et~al.}(2011)\citenamefont {H{\'e}bert-Dufresne}, \citenamefont {Allard},
  \citenamefont {Marceau}, \citenamefont {No{\"e}l},\ and\ \citenamefont
  {Dub{\'e}}}]{hebert2011}%
  \BibitemOpen
  \bibfield  {author} {\bibinfo {author} {\bibfnamefont {L.}~\bibnamefont
  {H{\'e}bert-Dufresne}}, \bibinfo {author} {\bibfnamefont {A.}~\bibnamefont
  {Allard}}, \bibinfo {author} {\bibfnamefont {V.}~\bibnamefont {Marceau}},
  \bibinfo {author} {\bibfnamefont {P.-A.}\ \bibnamefont {No{\"e}l}}, \ and\
  \bibinfo {author} {\bibfnamefont {L.~J.}\ \bibnamefont {Dub{\'e}}},\ }\href
  {\doibase 10.1103/PhysRevLett.107.158702} {\bibfield  {journal} {\bibinfo
  {journal} {Phys. Rev. Lett.}\ }\textbf {\bibinfo {volume} {107}},\ \bibinfo
  {pages} {158702} (\bibinfo {year} {2011})}\BibitemShut {NoStop}%
\bibitem [{\citenamefont {Zuev}\ \emph {et~al.}(2015)\citenamefont {Zuev},
  \citenamefont {Bogun{\'a}}, \citenamefont {Bianconi},\ and\ \citenamefont
  {Krioukov}}]{zuev2015}%
  \BibitemOpen
  \bibfield  {author} {\bibinfo {author} {\bibfnamefont {K.}~\bibnamefont
  {Zuev}}, \bibinfo {author} {\bibfnamefont {M.}~\bibnamefont {Bogun{\'a}}},
  \bibinfo {author} {\bibfnamefont {G.}~\bibnamefont {Bianconi}}, \ and\
  \bibinfo {author} {\bibfnamefont {D.}~\bibnamefont {Krioukov}},\ }\href@noop
  {} {\bibfield  {journal} {\bibinfo  {journal} {arXiv preprint
  arXiv:1501.06835}\ } (\bibinfo {year} {2015})}\BibitemShut {NoStop}%
\bibitem [{\citenamefont {Newman}(2001{\natexlab{b}})}]{newman2001i}%
  \BibitemOpen
  \bibfield  {author} {\bibinfo {author} {\bibfnamefont {M.~E.~J.}\
  \bibnamefont {Newman}},\ }\href {\doibase 10.1103/PhysRevE.64.016131}
  {\bibfield  {journal} {\bibinfo  {journal} {Phys. Rev. E}\ }\textbf {\bibinfo
  {volume} {64}},\ \bibinfo {pages} {016131} (\bibinfo {year}
  {2001}{\natexlab{b}})}\BibitemShut {NoStop}%
\bibitem [{\citenamefont {Newman}(2001{\natexlab{c}})}]{newman2001ii}%
  \BibitemOpen
  \bibfield  {author} {\bibinfo {author} {\bibfnamefont {M.~E.~J.}\
  \bibnamefont {Newman}},\ }\href {\doibase 10.1103/PhysRevE.64.016132}
  {\bibfield  {journal} {\bibinfo  {journal} {Phys. Rev. E}\ }\textbf {\bibinfo
  {volume} {64}},\ \bibinfo {pages} {016132} (\bibinfo {year}
  {2001}{\natexlab{c}})}\BibitemShut {NoStop}%
\bibitem [{\citenamefont {Olesen}\ \emph {et~al.}(2006)\citenamefont {Olesen},
  \citenamefont {Bascompte}, \citenamefont {Dupont},\ and\ \citenamefont
  {Jordano}}]{olesen2006}%
  \BibitemOpen
  \bibfield  {author} {\bibinfo {author} {\bibfnamefont {J.~M.}\ \bibnamefont
  {Olesen}}, \bibinfo {author} {\bibfnamefont {J.}~\bibnamefont {Bascompte}},
  \bibinfo {author} {\bibfnamefont {Y.~L.}\ \bibnamefont {Dupont}}, \ and\
  \bibinfo {author} {\bibfnamefont {P.}~\bibnamefont {Jordano}},\ }\href
  {\doibase http://dx.doi.org/10.1016/j.jtbi.2005.09.014} {\bibfield  {journal}
  {\bibinfo  {journal} {J. Theor. Biol.}\ }\textbf {\bibinfo {volume} {240}},\
  \bibinfo {pages} {270 } (\bibinfo {year} {2006})}\BibitemShut {NoStop}%
\bibitem [{\citenamefont {Poulin}(2010)}]{poulin2010}%
  \BibitemOpen
  \bibfield  {author} {\bibinfo {author} {\bibfnamefont {R.}~\bibnamefont
  {Poulin}},\ }\href {\doibase http://dx.doi.org/10.1016/j.pt.2010.05.008}
  {\bibfield  {journal} {\bibinfo  {journal} {Trends Parasitol.}\ }\textbf
  {\bibinfo {volume} {26}},\ \bibinfo {pages} {492 } (\bibinfo {year}
  {2010})}\BibitemShut {NoStop}%
\bibitem [{\citenamefont {Zhang}\ \emph {et~al.}(2015)\citenamefont {Zhang},
  \citenamefont {Dai}, \citenamefont {Li},\ and\ \citenamefont
  {Zhang}}]{zhang2015}%
  \BibitemOpen
  \bibfield  {author} {\bibinfo {author} {\bibfnamefont {D.}~\bibnamefont
  {Zhang}}, \bibinfo {author} {\bibfnamefont {M.}~\bibnamefont {Dai}}, \bibinfo
  {author} {\bibfnamefont {L.}~\bibnamefont {Li}}, \ and\ \bibinfo {author}
  {\bibfnamefont {C.}~\bibnamefont {Zhang}},\ }\href {\doibase
  http://dx.doi.org/10.1016/j.physa.2015.02.010} {\bibfield  {journal}
  {\bibinfo  {journal} {Physica A: Statistical Mechanics and its Applications}\
  }\textbf {\bibinfo {volume} {428}},\ \bibinfo {pages} {340 } (\bibinfo {year}
  {2015})}\BibitemShut {NoStop}%
\bibitem [{\citenamefont {Peruani}\ \emph {et~al.}(2007)\citenamefont
  {Peruani}, \citenamefont {Choudhury}, \citenamefont {Mukherjee},\ and\
  \citenamefont {Ganguly}}]{peruani2007}%
  \BibitemOpen
  \bibfield  {author} {\bibinfo {author} {\bibfnamefont {F.}~\bibnamefont
  {Peruani}}, \bibinfo {author} {\bibfnamefont {M.}~\bibnamefont {Choudhury}},
  \bibinfo {author} {\bibfnamefont {A.}~\bibnamefont {Mukherjee}}, \ and\
  \bibinfo {author} {\bibfnamefont {N.}~\bibnamefont {Ganguly}},\ }\href
  {http://stacks.iop.org/0295-5075/79/i=2/a=28001} {\bibfield  {journal}
  {\bibinfo  {journal} {EPL (Europhysics Letters)}\ }\textbf {\bibinfo {volume}
  {79}},\ \bibinfo {pages} {28001} (\bibinfo {year} {2007})}\BibitemShut
  {NoStop}%
\bibitem [{\citenamefont {Zhang}\ \emph {et~al.}(2013)\citenamefont {Zhang},
  \citenamefont {Zhang},\ and\ \citenamefont {Liu}}]{Zhang2013}%
  \BibitemOpen
  \bibfield  {author} {\bibinfo {author} {\bibfnamefont {C.-X.}\ \bibnamefont
  {Zhang}}, \bibinfo {author} {\bibfnamefont {Z.-K.}\ \bibnamefont {Zhang}}, \
  and\ \bibinfo {author} {\bibfnamefont {C.}~\bibnamefont {Liu}},\ }\href
  {\doibase http://dx.doi.org/10.1016/j.physa.2013.07.027} {\bibfield
  {journal} {\bibinfo  {journal} {Physica A: Statistical Mechanics and its
  Applications}\ }\textbf {\bibinfo {volume} {392}},\ \bibinfo {pages} {6100 }
  (\bibinfo {year} {2013})}\BibitemShut {NoStop}%
\bibitem [{\citenamefont {Bascompte}\ \emph {et~al.}(2003)\citenamefont
  {Bascompte}, \citenamefont {Jordano}, \citenamefont {Meli{\'a}n},\ and\
  \citenamefont {Olesen}}]{bascompte2003}%
  \BibitemOpen
  \bibfield  {author} {\bibinfo {author} {\bibfnamefont {J.}~\bibnamefont
  {Bascompte}}, \bibinfo {author} {\bibfnamefont {P.}~\bibnamefont {Jordano}},
  \bibinfo {author} {\bibfnamefont {C.~J.}\ \bibnamefont {Meli{\'a}n}}, \ and\
  \bibinfo {author} {\bibfnamefont {J.~M.}\ \bibnamefont {Olesen}},\ }\href
  {\doibase 10.1073/pnas.1633576100} {\bibfield  {journal} {\bibinfo  {journal}
  {Proc. Natl. Acad. Sci. U.S.A.}\ }\textbf {\bibinfo {volume} {100}},\
  \bibinfo {pages} {9383} (\bibinfo {year} {2003})}\BibitemShut {NoStop}%
\bibitem [{\citenamefont {Olesen}\ \emph {et~al.}(2007)\citenamefont {Olesen},
  \citenamefont {Bascompte}, \citenamefont {Dupont},\ and\ \citenamefont
  {Jordano}}]{olesen2007}%
  \BibitemOpen
  \bibfield  {author} {\bibinfo {author} {\bibfnamefont {J.~M.}\ \bibnamefont
  {Olesen}}, \bibinfo {author} {\bibfnamefont {J.}~\bibnamefont {Bascompte}},
  \bibinfo {author} {\bibfnamefont {Y.~L.}\ \bibnamefont {Dupont}}, \ and\
  \bibinfo {author} {\bibfnamefont {P.}~\bibnamefont {Jordano}},\ }\href
  {\doibase 10.1073/pnas.0706375104} {\bibfield  {journal} {\bibinfo  {journal}
  {Proc. Natl. Acad. Sci. U.S.A.}\ }\textbf {\bibinfo {volume} {104}},\
  \bibinfo {pages} {19891} (\bibinfo {year} {2007})},\ \Eprint
  {http://arxiv.org/abs/http://www.pnas.org/content/104/50/19891.full.pdf}
  {http://www.pnas.org/content/104/50/19891.full.pdf} \BibitemShut {NoStop}%
\bibitem [{\citenamefont {Fortuna}\ and\ \citenamefont
  {Bascompte}(2006)}]{fortuna2006}%
  \BibitemOpen
  \bibfield  {author} {\bibinfo {author} {\bibfnamefont {M.~A.}\ \bibnamefont
  {Fortuna}}\ and\ \bibinfo {author} {\bibfnamefont {J.}~\bibnamefont
  {Bascompte}},\ }\href {\doibase 10.1111/j.1461-0248.2005.00868.x} {\bibfield
  {journal} {\bibinfo  {journal} {Ecol. Lett.}\ }\textbf {\bibinfo {volume}
  {9}},\ \bibinfo {pages} {281} (\bibinfo {year} {2006})}\BibitemShut {NoStop}%
\bibitem [{\citenamefont {Th{\'e}bault}\ and\ \citenamefont
  {Fontaine}(2010)}]{thebault2010}%
  \BibitemOpen
  \bibfield  {author} {\bibinfo {author} {\bibfnamefont {E.}~\bibnamefont
  {Th{\'e}bault}}\ and\ \bibinfo {author} {\bibfnamefont {C.}~\bibnamefont
  {Fontaine}},\ }\href {\doibase 10.1126/science.1188321} {\bibfield  {journal}
  {\bibinfo  {journal} {Science}\ }\textbf {\bibinfo {volume} {329}},\ \bibinfo
  {pages} {853} (\bibinfo {year} {2010})},\ \Eprint
  {http://arxiv.org/abs/http://science.sciencemag.org/content/329/5993/853.full.pdf}
  {http://science.sciencemag.org/content/329/5993/853.full.pdf} \BibitemShut
  {NoStop}%
\bibitem [{\citenamefont {Stouffer}\ and\ \citenamefont
  {Bascompte}(2011)}]{stouffer2011}%
  \BibitemOpen
  \bibfield  {author} {\bibinfo {author} {\bibfnamefont {D.~B.}\ \bibnamefont
  {Stouffer}}\ and\ \bibinfo {author} {\bibfnamefont {J.}~\bibnamefont
  {Bascompte}},\ }\href {\doibase 10.1073/pnas.1014353108} {\bibfield
  {journal} {\bibinfo  {journal} {Proc. Natl. Acad. Sci. U.S.A.}\ }\textbf
  {\bibinfo {volume} {108}},\ \bibinfo {pages} {3648} (\bibinfo {year}
  {2011})},\ \Eprint
  {http://arxiv.org/abs/http://www.pnas.org/content/108/9/3648.full.pdf}
  {http://www.pnas.org/content/108/9/3648.full.pdf} \BibitemShut {NoStop}%
\bibitem [{\citenamefont {Fortuna}\ \emph {et~al.}(2010)\citenamefont
  {Fortuna}, \citenamefont {Stouffer}, \citenamefont {Olesen}, \citenamefont
  {Jordano}, \citenamefont {Mouillot}, \citenamefont {Krasnov}, \citenamefont
  {Poulin},\ and\ \citenamefont {Bascompte}}]{fortuna2010}%
  \BibitemOpen
  \bibfield  {author} {\bibinfo {author} {\bibfnamefont {M.~A.}\ \bibnamefont
  {Fortuna}}, \bibinfo {author} {\bibfnamefont {D.~B.}\ \bibnamefont
  {Stouffer}}, \bibinfo {author} {\bibfnamefont {J.~M.}\ \bibnamefont
  {Olesen}}, \bibinfo {author} {\bibfnamefont {P.}~\bibnamefont {Jordano}},
  \bibinfo {author} {\bibfnamefont {D.}~\bibnamefont {Mouillot}}, \bibinfo
  {author} {\bibfnamefont {B.~R.}\ \bibnamefont {Krasnov}}, \bibinfo {author}
  {\bibfnamefont {R.}~\bibnamefont {Poulin}}, \ and\ \bibinfo {author}
  {\bibfnamefont {J.}~\bibnamefont {Bascompte}},\ }\href {\doibase
  10.1111/j.1365-2656.2010.01688.x} {\bibfield  {journal} {\bibinfo  {journal}
  {J. Anim. Ecol.}\ }\textbf {\bibinfo {volume} {79}},\ \bibinfo {pages} {811}
  (\bibinfo {year} {2010})}\BibitemShut {NoStop}%
\bibitem [{\citenamefont {Jover}\ \emph {et~al.}(2015)\citenamefont {Jover},
  \citenamefont {Flores}, \citenamefont {Cortez},\ and\ \citenamefont
  {Weitz}}]{jover2015}%
  \BibitemOpen
  \bibfield  {author} {\bibinfo {author} {\bibfnamefont {L.~F.}\ \bibnamefont
  {Jover}}, \bibinfo {author} {\bibfnamefont {C.~O.}\ \bibnamefont {Flores}},
  \bibinfo {author} {\bibfnamefont {M.~H.}\ \bibnamefont {Cortez}}, \ and\
  \bibinfo {author} {\bibfnamefont {J.~S.}\ \bibnamefont {Weitz}},\ }\href@noop
  {} {\bibfield  {journal} {\bibinfo  {journal} {Sci. Rep.}\ }\textbf {\bibinfo
  {volume} {5}},\ \bibinfo {pages} {17856} (\bibinfo {year}
  {2015})}\BibitemShut {NoStop}%
\bibitem [{\citenamefont {Debar}\ \emph {et~al.}(1999)\citenamefont {Debar},
  \citenamefont {Dacier},\ and\ \citenamefont {Wespi}}]{debar_1999}%
  \BibitemOpen
  \bibfield  {author} {\bibinfo {author} {\bibfnamefont {H.}~\bibnamefont
  {Debar}}, \bibinfo {author} {\bibfnamefont {M.}~\bibnamefont {Dacier}}, \
  and\ \bibinfo {author} {\bibfnamefont {A.}~\bibnamefont {Wespi}},\ }\href
  {\doibase 10.1016/S1389-1286(98)00017-6} {\bibfield  {journal} {\bibinfo
  {journal} {Comput. Netw.}\ }\textbf {\bibinfo {volume} {31}},\ \bibinfo
  {pages} {805 } (\bibinfo {year} {1999})}\BibitemShut {NoStop}%
\bibitem [{\citenamefont {Krebs}(2002)}]{krebs_2002}%
  \BibitemOpen
  \bibfield  {author} {\bibinfo {author} {\bibfnamefont {V.~E.}\ \bibnamefont
  {Krebs}},\ }\href@noop {} {\bibfield  {journal} {\bibinfo  {journal}
  {Connections}\ }\textbf {\bibinfo {volume} {24}},\ \bibinfo {pages} {43}
  (\bibinfo {year} {2002})}\BibitemShut {NoStop}%
\bibitem [{\citenamefont {Borgatti}\ \emph {et~al.}(2009)\citenamefont
  {Borgatti}, \citenamefont {Mehra}, \citenamefont {Brass},\ and\ \citenamefont
  {Labianca}}]{borgatti_2009}%
  \BibitemOpen
  \bibfield  {author} {\bibinfo {author} {\bibfnamefont {S.~P.}\ \bibnamefont
  {Borgatti}}, \bibinfo {author} {\bibfnamefont {A.}~\bibnamefont {Mehra}},
  \bibinfo {author} {\bibfnamefont {D.~J.}\ \bibnamefont {Brass}}, \ and\
  \bibinfo {author} {\bibfnamefont {G.}~\bibnamefont {Labianca}},\ }\href
  {\doibase 10.1126/science.1165821} {\bibfield  {journal} {\bibinfo  {journal}
  {Science}\ }\textbf {\bibinfo {volume} {323}},\ \bibinfo {pages} {892}
  (\bibinfo {year} {2009})}\BibitemShut {NoStop}%
\bibitem [{\citenamefont {Feiner}\ \emph {et~al.}(2015)\citenamefont {Feiner},
  \citenamefont {Argov}, \citenamefont {Rabinovich}, \citenamefont {Sigal},
  \citenamefont {Borovok},\ and\ \citenamefont {Herskovits}}]{feiner2015}%
  \BibitemOpen
  \bibfield  {author} {\bibinfo {author} {\bibfnamefont {R.}~\bibnamefont
  {Feiner}}, \bibinfo {author} {\bibfnamefont {T.}~\bibnamefont {Argov}},
  \bibinfo {author} {\bibfnamefont {L.}~\bibnamefont {Rabinovich}}, \bibinfo
  {author} {\bibfnamefont {N.}~\bibnamefont {Sigal}}, \bibinfo {author}
  {\bibfnamefont {I.}~\bibnamefont {Borovok}}, \ and\ \bibinfo {author}
  {\bibfnamefont {A.~A.}\ \bibnamefont {Herskovits}},\ }\href@noop {}
  {\bibfield  {journal} {\bibinfo  {journal} {Nature Rev. Microbiol.}\ }\textbf
  {\bibinfo {volume} {13}},\ \bibinfo {pages} {641} (\bibinfo {year}
  {2015})}\BibitemShut {NoStop}%
\bibitem [{\citenamefont {Luria}\ and\ \citenamefont
  {Delbr\"{u}ck}(1943)}]{luria_1943}%
  \BibitemOpen
  \bibfield  {author} {\bibinfo {author} {\bibfnamefont {S.~E.}\ \bibnamefont
  {Luria}}\ and\ \bibinfo {author} {\bibfnamefont {M.}~\bibnamefont
  {Delbr\"{u}ck}},\ }\href@noop {} {\bibfield  {journal} {\bibinfo  {journal}
  {Genetics}\ }\textbf {\bibinfo {volume} {28}},\ \bibinfo {pages} {491}
  (\bibinfo {year} {1943})}\BibitemShut {NoStop}%
\bibitem [{\citenamefont {Luria}(1945)}]{luria_1945}%
  \BibitemOpen
  \bibfield  {author} {\bibinfo {author} {\bibfnamefont {S.~E.}\ \bibnamefont
  {Luria}},\ }\href@noop {} {\bibfield  {journal} {\bibinfo  {journal}
  {Genetics}\ }\textbf {\bibinfo {volume} {30}},\ \bibinfo {pages} {84}
  (\bibinfo {year} {1945})}\BibitemShut {NoStop}%
\bibitem [{\citenamefont {Weitz}\ \emph {et~al.}(2013)\citenamefont {Weitz},
  \citenamefont {Poisot}, \citenamefont {Meyer}, \citenamefont {Flores},
  \citenamefont {Valverde}, \citenamefont {Sullivan},\ and\ \citenamefont
  {Hochberg}}]{weitz2013}%
  \BibitemOpen
  \bibfield  {author} {\bibinfo {author} {\bibfnamefont {J.~S.}\ \bibnamefont
  {Weitz}}, \bibinfo {author} {\bibfnamefont {T.}~\bibnamefont {Poisot}},
  \bibinfo {author} {\bibfnamefont {J.~R.}\ \bibnamefont {Meyer}}, \bibinfo
  {author} {\bibfnamefont {C.~O.}\ \bibnamefont {Flores}}, \bibinfo {author}
  {\bibfnamefont {S.}~\bibnamefont {Valverde}}, \bibinfo {author}
  {\bibfnamefont {M.~B.}\ \bibnamefont {Sullivan}}, \ and\ \bibinfo {author}
  {\bibfnamefont {M.~E.}\ \bibnamefont {Hochberg}},\ }\href {\doibase
  10.1016/j.tim.2012.11.003} {\bibfield  {journal} {\bibinfo  {journal} {Trends
  Microbiol.}\ }\textbf {\bibinfo {volume} {21}},\ \bibinfo {pages} {82}
  (\bibinfo {year} {2013})}\BibitemShut {NoStop}%
\bibitem [{\citenamefont {Flores}\ \emph {et~al.}(2011)\citenamefont {Flores},
  \citenamefont {Meyer}, \citenamefont {Valverde}, \citenamefont {Farr},\ and\
  \citenamefont {Weitz}}]{flores2011}%
  \BibitemOpen
  \bibfield  {author} {\bibinfo {author} {\bibfnamefont {C.~O.}\ \bibnamefont
  {Flores}}, \bibinfo {author} {\bibfnamefont {J.~R.}\ \bibnamefont {Meyer}},
  \bibinfo {author} {\bibfnamefont {S.}~\bibnamefont {Valverde}}, \bibinfo
  {author} {\bibfnamefont {L.}~\bibnamefont {Farr}}, \ and\ \bibinfo {author}
  {\bibfnamefont {J.~S.}\ \bibnamefont {Weitz}},\ }\href {\doibase
  10.1073/pnas.1101595108} {\bibfield  {journal} {\bibinfo  {journal} {Proc.
  Natl. Acad. Sci. U.S.A.}\ }\textbf {\bibinfo {volume} {108}},\ \bibinfo
  {pages} {E288} (\bibinfo {year} {2011})}\BibitemShut {NoStop}%
\bibitem [{\citenamefont {Flores}\ \emph {et~al.}(2013)\citenamefont {Flores},
  \citenamefont {Valverde},\ and\ \citenamefont {Weitz}}]{flores2013}%
  \BibitemOpen
  \bibfield  {author} {\bibinfo {author} {\bibfnamefont {C.~O.}\ \bibnamefont
  {Flores}}, \bibinfo {author} {\bibfnamefont {S.}~\bibnamefont {Valverde}}, \
  and\ \bibinfo {author} {\bibfnamefont {J.~S.}\ \bibnamefont {Weitz}},\ }\href
  {\doibase 10.1038/ismej.2012.135} {\bibfield  {journal} {\bibinfo  {journal}
  {ISME J.}\ }\textbf {\bibinfo {volume} {7}},\ \bibinfo {pages} {520}
  (\bibinfo {year} {2013})}\BibitemShut {NoStop}%
\bibitem [{\citenamefont {Poullain}\ \emph {et~al.}(2008)\citenamefont
  {Poullain}, \citenamefont {Gandon}, \citenamefont {Brockhurst}, \citenamefont
  {Buckling},\ and\ \citenamefont {Hochberg}}]{poullain_2008}%
  \BibitemOpen
  \bibfield  {author} {\bibinfo {author} {\bibfnamefont {V.}~\bibnamefont
  {Poullain}}, \bibinfo {author} {\bibfnamefont {S.}~\bibnamefont {Gandon}},
  \bibinfo {author} {\bibfnamefont {M.~A.}\ \bibnamefont {Brockhurst}},
  \bibinfo {author} {\bibfnamefont {A.}~\bibnamefont {Buckling}}, \ and\
  \bibinfo {author} {\bibfnamefont {M.~E.}\ \bibnamefont {Hochberg}},\ }\href
  {\doibase 10.1111/j.1558-5646.2007.00260.x} {\bibfield  {journal} {\bibinfo
  {journal} {Evolution}\ }\textbf {\bibinfo {volume} {62}},\ \bibinfo {pages}
  {1} (\bibinfo {year} {2008})}\BibitemShut {NoStop}%
\bibitem [{\citenamefont {Scanlan}\ \emph {et~al.}(2011)\citenamefont
  {Scanlan}, \citenamefont {Hall}, \citenamefont {Lopez-Pascua},\ and\
  \citenamefont {Buckling}}]{scanlan2011}%
  \BibitemOpen
  \bibfield  {author} {\bibinfo {author} {\bibfnamefont {P.~D.}\ \bibnamefont
  {Scanlan}}, \bibinfo {author} {\bibfnamefont {A.~R.}\ \bibnamefont {Hall}},
  \bibinfo {author} {\bibfnamefont {L.~D.}\ \bibnamefont {Lopez-Pascua}}, \
  and\ \bibinfo {author} {\bibfnamefont {A.}~\bibnamefont {Buckling}},\ }\href
  {\doibase 10.1111/j.1365-294X.2010.04903.x} {\bibfield  {journal} {\bibinfo
  {journal} {Mol. Ecol.}\ }\textbf {\bibinfo {volume} {20}},\ \bibinfo {pages}
  {981} (\bibinfo {year} {2011})}\BibitemShut {NoStop}%
\bibitem [{\citenamefont {Meyer}\ \emph {et~al.}(2012)\citenamefont {Meyer},
  \citenamefont {Dobias}, \citenamefont {Weitz}, \citenamefont {Barrick},
  \citenamefont {Quick},\ and\ \citenamefont
  {Lenski}}]{meyer_2012repeatability}%
  \BibitemOpen
  \bibfield  {author} {\bibinfo {author} {\bibfnamefont {J.~R.}\ \bibnamefont
  {Meyer}}, \bibinfo {author} {\bibfnamefont {D.~T.}\ \bibnamefont {Dobias}},
  \bibinfo {author} {\bibfnamefont {J.~S.}\ \bibnamefont {Weitz}}, \bibinfo
  {author} {\bibfnamefont {J.~E.}\ \bibnamefont {Barrick}}, \bibinfo {author}
  {\bibfnamefont {R.~T.}\ \bibnamefont {Quick}}, \ and\ \bibinfo {author}
  {\bibfnamefont {R.~E.}\ \bibnamefont {Lenski}},\ }\href {\doibase
  10.1126/science.1214449} {\bibfield  {journal} {\bibinfo  {journal}
  {Science}\ }\textbf {\bibinfo {volume} {335}},\ \bibinfo {pages} {428}
  (\bibinfo {year} {2012})}\BibitemShut {NoStop}%
\bibitem [{\citenamefont {Brockhurst}\ and\ \citenamefont
  {Koskella}(2013)}]{brockhurst2013}%
  \BibitemOpen
  \bibfield  {author} {\bibinfo {author} {\bibfnamefont {M.~A.}\ \bibnamefont
  {Brockhurst}}\ and\ \bibinfo {author} {\bibfnamefont {B.}~\bibnamefont
  {Koskella}},\ }\href {\doibase http://dx.doi.org/10.1016/j.tree.2013.02.009}
  {\bibfield  {journal} {\bibinfo  {journal} {Trends Ecol. Evol.}\ }\textbf
  {\bibinfo {volume} {28}},\ \bibinfo {pages} {367 } (\bibinfo {year}
  {2013})}\BibitemShut {NoStop}%
\bibitem [{\citenamefont {Sieber}\ \emph {et~al.}(2014)\citenamefont {Sieber},
  \citenamefont {Robb}, \citenamefont {Forde},\ and\ \citenamefont
  {Gudelj}}]{sieber_isme2014}%
  \BibitemOpen
  \bibfield  {author} {\bibinfo {author} {\bibfnamefont {M.}~\bibnamefont
  {Sieber}}, \bibinfo {author} {\bibfnamefont {M.}~\bibnamefont {Robb}},
  \bibinfo {author} {\bibfnamefont {S.~E.}\ \bibnamefont {Forde}}, \ and\
  \bibinfo {author} {\bibfnamefont {I.}~\bibnamefont {Gudelj}},\ }\href
  {\doibase 10.1038/ismej.2013.169} {\bibfield  {journal} {\bibinfo  {journal}
  {ISME J.}\ }\textbf {\bibinfo {volume} {8}},\ \bibinfo {pages} {504}
  (\bibinfo {year} {2014})}\BibitemShut {NoStop}%
\bibitem [{\citenamefont {Newman}(2006)}]{newman2006}%
  \BibitemOpen
  \bibfield  {author} {\bibinfo {author} {\bibfnamefont {M.~E.~J.}\
  \bibnamefont {Newman}},\ }\href {\doibase 10.1073/pnas.0601602103} {\bibfield
   {journal} {\bibinfo  {journal} {Proc. Natl. Acad. Sci. U.S.A.}\ }\textbf
  {\bibinfo {volume} {103}},\ \bibinfo {pages} {8577} (\bibinfo {year}
  {2006})}\BibitemShut {NoStop}%
\bibitem [{\citenamefont {Barber}(2007)}]{barber2007}%
  \BibitemOpen
  \bibfield  {author} {\bibinfo {author} {\bibfnamefont {M.~J.}\ \bibnamefont
  {Barber}},\ }\href {\doibase 10.1103/PhysRevE.76.066102} {\bibfield
  {journal} {\bibinfo  {journal} {Phys. Rev. E}\ }\textbf {\bibinfo {volume}
  {76}},\ \bibinfo {pages} {066102} (\bibinfo {year} {2007})}\BibitemShut
  {NoStop}%
\bibitem [{\citenamefont {Almeida-Neto}\ \emph {et~al.}(2008)\citenamefont
  {Almeida-Neto}, \citenamefont {Guimaraes}, \citenamefont {Guimar{\~a}es},
  \citenamefont {Loyola},\ and\ \citenamefont {Ulrich}}]{almeida2008}%
  \BibitemOpen
  \bibfield  {author} {\bibinfo {author} {\bibfnamefont {M.}~\bibnamefont
  {Almeida-Neto}}, \bibinfo {author} {\bibfnamefont {P.}~\bibnamefont
  {Guimaraes}}, \bibinfo {author} {\bibfnamefont {P.~R.}\ \bibnamefont
  {Guimar{\~a}es}}, \bibinfo {author} {\bibfnamefont {R.~D.}\ \bibnamefont
  {Loyola}}, \ and\ \bibinfo {author} {\bibfnamefont {W.}~\bibnamefont
  {Ulrich}},\ }\href {\doibase 10.1111/j.0030-1299.2008.16644.x} {\bibfield
  {journal} {\bibinfo  {journal} {Oikos}\ }\textbf {\bibinfo {volume} {117}},\
  \bibinfo {pages} {1227} (\bibinfo {year} {2008})}\BibitemShut {NoStop}%
\bibitem [{\citenamefont {Flores}\ \emph {et~al.}(2014)\citenamefont {Flores},
  \citenamefont {Poisot},\ and\ \citenamefont {Weitz}}]{flores2014}%
  \BibitemOpen
  \bibfield  {author} {\bibinfo {author} {\bibfnamefont {C.~O.}\ \bibnamefont
  {Flores}}, \bibinfo {author} {\bibfnamefont {T.}~\bibnamefont {Poisot}}, \
  and\ \bibinfo {author} {\bibfnamefont {J.~S.}\ \bibnamefont {Weitz}},\
  }\href@noop {} {\bibfield  {journal} {\bibinfo  {journal} {arXiv preprint
  arXiv:1406.6732}\ } (\bibinfo {year} {2014})}\BibitemShut {NoStop}%
\bibitem [{\citenamefont {Jover}\ \emph {et~al.}(2013)\citenamefont {Jover},
  \citenamefont {Cortez},\ and\ \citenamefont {Weitz}}]{jover_jtb2013}%
  \BibitemOpen
  \bibfield  {author} {\bibinfo {author} {\bibfnamefont {L.~F.}\ \bibnamefont
  {Jover}}, \bibinfo {author} {\bibfnamefont {M.~H.}\ \bibnamefont {Cortez}}, \
  and\ \bibinfo {author} {\bibfnamefont {J.~S.}\ \bibnamefont {Weitz}},\ }\href
  {\doibase 10.1016/j.jtbi.2013.04.011} {\bibfield  {journal} {\bibinfo
  {journal} {J. Theor. Biol.}\ }\textbf {\bibinfo {volume} {332}},\ \bibinfo
  {pages} {65} (\bibinfo {year} {2013})}\BibitemShut {NoStop}%
\bibitem [{\citenamefont {Korytowski}\ and\ \citenamefont
  {Smith}(2015)}]{korytowski_2015}%
  \BibitemOpen
  \bibfield  {author} {\bibinfo {author} {\bibfnamefont {D.~A.}\ \bibnamefont
  {Korytowski}}\ and\ \bibinfo {author} {\bibfnamefont {H.~L.}\ \bibnamefont
  {Smith}},\ }\href {\doibase 10.1007/s12080-014-0236-6} {\bibfield  {journal}
  {\bibinfo  {journal} {Theor. Ecol.}\ }\textbf {\bibinfo {volume} {8}},\
  \bibinfo {pages} {111} (\bibinfo {year} {2015})}\BibitemShut {NoStop}%
\bibitem [{\citenamefont {Thingstad}(2000)}]{thingstad2000}%
  \BibitemOpen
  \bibfield  {author} {\bibinfo {author} {\bibfnamefont {T.~F.}\ \bibnamefont
  {Thingstad}},\ }\href {\doibase 10.4319/lo.2000.45.6.1320} {\bibfield
  {journal} {\bibinfo  {journal} {Limnol. Oceanogr.}\ }\textbf {\bibinfo
  {volume} {45}},\ \bibinfo {pages} {1320} (\bibinfo {year}
  {2000})}\BibitemShut {NoStop}%
\bibitem [{\citenamefont {Winter}\ \emph {et~al.}(2010)\citenamefont {Winter},
  \citenamefont {Bouvier}, \citenamefont {Weinbauer},\ and\ \citenamefont
  {Thingstad}}]{winter2010}%
  \BibitemOpen
  \bibfield  {author} {\bibinfo {author} {\bibfnamefont {C.}~\bibnamefont
  {Winter}}, \bibinfo {author} {\bibfnamefont {T.}~\bibnamefont {Bouvier}},
  \bibinfo {author} {\bibfnamefont {M.~G.}\ \bibnamefont {Weinbauer}}, \ and\
  \bibinfo {author} {\bibfnamefont {T.~F.}\ \bibnamefont {Thingstad}},\ }\href
  {\doibase 10.1128/MMBR.00034-09} {\bibfield  {journal} {\bibinfo  {journal}
  {Microbiol. Mol. Biol. Rev.}\ }\textbf {\bibinfo {volume} {74}},\ \bibinfo
  {pages} {42} (\bibinfo {year} {2010})}\BibitemShut {NoStop}%
\bibitem [{\citenamefont {Rosvall}\ \emph {et~al.}(2006)\citenamefont
  {Rosvall}, \citenamefont {Dodd}, \citenamefont {Krishna},\ and\ \citenamefont
  {Sneppen}}]{rosvall2006}%
  \BibitemOpen
  \bibfield  {author} {\bibinfo {author} {\bibfnamefont {M.}~\bibnamefont
  {Rosvall}}, \bibinfo {author} {\bibfnamefont {I.~B.}\ \bibnamefont {Dodd}},
  \bibinfo {author} {\bibfnamefont {S.}~\bibnamefont {Krishna}}, \ and\
  \bibinfo {author} {\bibfnamefont {K.}~\bibnamefont {Sneppen}},\ }\href
  {\doibase 10.1103/PhysRevE.74.066105} {\bibfield  {journal} {\bibinfo
  {journal} {Phys. Rev. E}\ }\textbf {\bibinfo {volume} {74}},\ \bibinfo
  {pages} {066105} (\bibinfo {year} {2006})}\BibitemShut {NoStop}%
\end{thebibliography}%
\bibliographystyle{apsrev4-1}

\end{document}